\documentclass[epj]{svjour}
\usepackage{epsfig,array}
\usepackage{multirow}
\usepackage{graphicx}
\newcommand{\be}{\begin{eqnarray}}
\newcommand{\ee}{\end{eqnarray}}

\newcommand{\nn}{\nonumber }
\newcommand{\bea}{\begin{eqnarray}}
\newcommand{\eea}{\end{eqnarray}}

\newcommand{\bg}{\mbox{\boldmath$\gamma$}}
\newcommand{\bs}{\mbox{\boldmath$\sigma$}}
\newcommand{\bta}{\mbox{\boldmath$\tau$}}

\newcommand{\beq}{\begin{equation}}
\newcommand{\eeq}{\end{equation}}

\def\la{\mathrel{\mathpalette\fun <}}

\def\fun#1#2{\lower3.6pt\vbox{\baselineskip0pt\lineskip.9pt
\ialign{$\mathsurround=0pt#1\hfil##\hfil$\crcr#2\crcr\sim\crcr}}}

\begin{document}

\title{Baryon-baryon and baryon-antibaryon interaction amplitudes in the
spin--momentum operator expansion method}

\author{
     A.V.~Anisovich \inst{1,2} \and V.V.~Anisovich \inst{2}
\and E.~Klempt \inst{1}
\and V.A.~Nikonov   \inst{1,2}
\and A.V.~Sarantsev \inst{1,2}
}

\institute{
Helmholtz--Institut f\"ur Strahlen-- und Kernphysik,
Universit\"at Bonn, Germany
\and Petersburg Nuclear Physics Institute, Gatchina, 188300 Russia
}

\date{Received: \today / }

\abstract{Partial wave scattering amplitudes in baryon-baryon and baryon-antibaryon
collisions and amplitudes for the production and decay of baryon
resonances are constructed in the framework of the spin-momentum
operator expansion method. The approach is relativistically invariant
and it allows us to perform combined analyses of different reactions
imposing analyticity and unitarity directly. The role of final state
interactions (triangle and box diagrams) is discussed.
 }
\authorrunning{A.V.~Anisovich {\it et al.}}
\titlerunning{Baryon-baryon and baryon-antibaryon interaction ...}
\mail{nikonov@iskp.uni-bonn.de}

\maketitle

\section{Introduction}

To understand strong interactions at low and intermediate
energies is one of the important tasks when quantum chromodynamics is
being studied. At large momentum transfer, QCD can be used efficiently
due to the smallness of the strong interaction coupling constant
\cite{Gross:1973id}; the low-energy domain can be treated using
effective field theories \cite{Weinberg:1991um}. The resonance region
is much more difficult to access. Lattice gauge calculations are
capable to reproduce the masses of ground state hadrons
\cite{Aoki:1999yr} but excited states and their decay properties are
difficult to extract from lattice data. For further progress,
systematic experimental information seems to be mandatory to identify
the leading mechanisms responsible for the mass spectrum and for the
decay amplitudes of strongly interacting particles.

Recently, considerable progress has been achieved in meson
spectroscopy, even though a commonly agreed picture has not yet
emerged. Recent reviews emphasizing different views can be found in
\cite{Amsler:2004ps,Bugg:2004xu,Anisovich:2005jn,Klempt:2007}. The main sources of
recent progress were the study of reactions with multi-particle final
states. The analysis of data on proton-anti\-proton annihilation at rest
resulted in the discovery of a number of particles in the region
1300-1800 MeV \cite{Amsler:1994rv}-\cite{Amsler:1995bz}; the
investigation of the proton-antiproton annihilation in flight led to a
large set of new states over the region 1800-2500 MeV
\cite{Anisovich:2000ut}-\cite{Anisovich:2002sv}. It appeared that the
majority of the newly discovered states are lying on linear trajectories
against radial excitation number \cite{Anisovich:2000kx}. Such a
pattern was not predicted by the classical quark model of Godfrey and
Isgur \cite{Godfrey:1985xj} using a linear confinement poten\-tial and
additional interactions due to effective one-gluon exchange forces.
More recent calculations based on instanton-indu\-ced interactions
\cite{Loring:2001kx,Loring:2001ky}
 can,
however, be tuned (by choosing an appropriate Dirac structure of the
confinement potential) to reproduce the observed mass pat\-tern very
well. The pattern can be understood, too, with\-in a 5-dimensional
theory holographically dual to QCD (AdS/ QCD) \cite{Karch:2006pv} which
predicts masses to proportional to $(N+L)$ where $L$ is the intrinsic
orbital angular momentum between quark and antiquark and $N$ a radial
quantum number. In the light-quark meson spectrum, practically all
expected states are observed. However, there are a few additional
states which do not belong to these trajectories. These states are
candidates to be of exotic nature, e.g., they could be glueballs or
hybrids.

The situation in the baryon sector is in some sense reverse:  for baryons, the
quark model predicts a much larger number of states than that observed
experimentally. So far, the pattern seems to suggest that not all degrees of
freedom in the three-quark system are realised in the spectrum of excited
states. Instead, the pattern of excited states follows the same $(L+N)$ pattern
\cite{Klempt:2002vp} which is observed for mesons. If this is the case, the
fact would be an important phenomenon in the physics of highly excited states.
Still, a detailed verification of this statement is needed. On the other hand,
the main information on baryon resonances has come from the $\pi N$ elastic
scattering, and one may hope that many new states will be discovered in (i)
reactions involving strangeness in two hadron final states and in (ii)
inelastic reactions induced by photons or protons with three or four particles
in the final states (for example, two pion photoproduction).

The search for new baryon resonances is of topical interest and several
experiments like  CB-ELSA, CLAS, GRAAL, SAPHIR, and SPRING-8
pursue active searches using photoproduction as a tool
\cite{Bartholomy:04,Crede:04,Krusche:nv,GRAAL1,SAID1,Beck,Reb,GRAAL2,%
SAID2,Dur,Buch,Glander:2003jw,McNabb:2003nf,Zegers:2003ux,Lawall:2005np,%
Braghieri:1994rf,Harter:1997jq,Wolf:2000qt,Ahrens:2005ia,%
Assafiri:2003mv,Ripani:2002ss,Strauch:2005cs,Wu:2005wf,FT,%
Barth03,Hourany:2005wh,Bradford:2006ba,Lleres:2007tx}.
 A few new resonances were suggested
\cite{our_fit1,our_fit2,our_fit3} in fits to these data sets.
Proton-proton collision experiments can $\;$provide an important source
of information on baryon resonances including exotic states (e.g.
pentaquarks \cite{Diakonov:1997mm,Jaffe:2003sg,Nakano:2003qx,%
Stepanyan:2003qr,Barmin:2003vv,Barth:2003es,Abdel-Bary:2004ts,%
Abdel-Bary:2006wd}). COSY at the Research Center J\"ulich is providing
a wealth of data on meson production in proton-proton inelastic
scattering \cite{Balewski:1991ns,Bondar:1995zv,Calen:1996mn,%
Calen:1997sf,Zlomanczuk:1998fh,Betsch:1998ir,Calen:1998kx,%
Sewerin:1998ky,Calen:1998mg,Moskal:1998pc,Balewski:1998pd,Calen:1998vh,%
Moskal:2000gj,Balestra:2000ic,Greiff:2000nn,Moskal:2000pu,%
AbdEl-Samad:2001dv,Quentmeier:2001ec,Moskal:2001gm,Bilger:2001qz,%
Brodowski:2002fg,Winter:2002ft,Greiff:2002fv,Moskal:2002jm,%
Abdel-Bary:2002sx,Brodowski:2002xw,Kondratyuk:2002yf,Moskal:2003gt,%
Abdel-Bary:2003gz,Kleber:2003kx,Patzold:2003tr,Yaschenko:2004az,%
Abdel-Bary:2004qb,Grishina:2004rd,Abdel-Bary:2005bj,Kleber:2005bz,%
Wronska:2005dt,Valdau:2005ed,Dymov:2005nc,Zychor:2005sj,Moskal:2005uh,%
Abdel-Bary:2005zb,Dzyuba:2006bj,Abdel-Bary:2006,Rozek:2006ct,%
Barsov:2006sc,Winter:2006vd,Maeda:2006wv,El-Samad:2006wy,Uzikov:2006yc,%
Hartmann:2006zc,Abdel-Bary:2007sq}. The experiments Anke and COSY-11
cove\-red mainly the threshold region, while TOF covers the full
dynamical range. At present, the upgraded WASA detector is installed at
COSY and will provide high-statistics data on the production of neutral
mesons in $NN$ interactions \cite{Adam:2004ch}. The data provide
stringent information on nucleon--nucleon--meson vertices and on the
formation of baryon resonances. Selected papers can be found in
\cite{Hanhart:1997jd,Hanhart:1998rn,Bernard:1998sz,Alvarez-Ruso:1998xg,Hanhart:1998za,%
Niskanen:1998zf,Dmitrasinovic:1999cu,Kaiser:1999cs,Kaiser:1999ia,%
Nakayama:1999jb,Machner:1999ky,Shyam:1999nm,Shyam:2000sn,Sibirtsev:2000ut,%
Hanhart:2002bu,Nakayama:2002mu,Baru:2002rs,Nakayama:2003jn,Hanhart:2003pg,%
Deloff:2003te,Hanhart:2003xp,Nakayama:2004ek,Hanhart:2004re,Baru:2004xg,%
Lensky:2005jc,Shyam:2005dw,Schneider:2006bd}.

The partial wave analysis of such processes cannot be carried out without
taking into account the final state interaction. In many processes, the
inclusion of the proton-proton interaction dramatically changes the
description of the data \cite{Sibirtsev:2005mv}. However, a number of
important problems for such analyses has not comprehensively developed
yet. Among these problems are a correct treatment of relativistic
effects and of the contributions of triangle or box diagrams.

In this paper, we present a relativistically invariant approach for the
partial wave analysis of proton-proton interactions. The method is
based on the spin-momentum operator expansion  suggested in
\cite{xoper,ASB,decompos,Anisovich:2006bc}. The contribution of
triangle and box diagrams to the meson production processes is
discussed and certain examples are considered.

In Section 2, we present the partial wave expansion for baryon-baryon
and baryon-antibaryon scattering amplitudes. In  Section 3, the
unitarity condition for fermion-fermion partial wave amplitudes is
discussed. The angular momentum and spin operators for nucleon-nucleon
scattering are introduced in Section 4, the nucleon-nucleon partial
wave amplitude is constructed in Section 5. In this Section,
fermion-fermion one-loop diagrams and the cross section for the
two-fermion scattering are calculated. The operators for $N\Delta$
production are constructed in the Section 6. Some examples of
amplitudes with multi-particle final states are given in Section 7.
Properties of the triangle and box diagrams are shortly
discussed in Sections 8 and 9.

\section{Selection rules for baryon-antibaryon and baryon-baryon
scattering amplitudes}

\subsection{Baryon pairs with isospin \boldmath $I=0$}

First, consider the baryon--antibaryon scattering amplitude in a
isospin singlet configuration, for example, the $\Lambda\bar\Lambda$
scattering amplitude. One can use two alternative representations of
the baryon-antibaryon amplitude
$\Lambda(p_1)\bar\Lambda(p_2)$ $\to$ $\Lambda(p'_1)\bar\Lambda(p'_2)$.

In the $t$-channel representation the amplitude is the sum of partial
waves in the $t$-channel with definite quantum numbers: spin $S$,
angular momentum $L$ and total momentum $J$ (we define
$t=q^2=(p'_1-p_1)^2$):
\be \label{TB14}
&&M(s,t,u)= \sum_{S,L,L',J\atop
\mu_1\ldots\mu_J}\left(\bar\psi(p'_1) \tilde Q^{SLJ}_{\mu_1\ldots\mu_J}
(q)\psi(p_1)\right)
\nonumber \\
&&\times
\left(
\bar\psi(p'_2)\tilde Q^{SL'J}_{\mu_1\ldots\mu_J} (q)
\psi(p_2)\right)A^{(S,L'L,J)}_t(q^2).
\ee
Here, $\tilde Q$ is the $t$-channel operator (the four-component
spi\-nors $\psi(p)$ are given in Appendix 1) and
$\mu_1,\mu_2\ldots\mu_{J-1},\mu_J$ are the indices of the rank $J$ operator.

Another representation is related to the $s$-channel (we define
$s=(p_1+p_2)^2$):
\be \label{TB16}
&&M(s,t,u)= \sum_{S,L,L',J \atop
\mu_1\ldots\mu_J} \left(\bar\psi (p'_1)
Q^{SL'J}_{\mu_1\ldots\mu_J}(k'_\perp)\psi^c(-p'_2)\right)
\nonumber \\
&&\times
\left(\bar\psi^c(-p_2)
Q^{SLJ}_{\mu_1\ldots\mu_J}(k_\perp)\psi(p_1)\right)
A^{(S,L'L,J)}_s(s)\, .
\ee
Here, $\psi^c(-p)$ are charge conjugated four-component
spi\-nors (see Appendix 1)
and $Q^{SLJ}_{\mu_1\ldots\mu_J}$ are the $s$-channel operators, where
$S,L,J$ are,
correspondingly, spin, angular momentum and total momentum of the
partial wave in the $s$-channel. The notations of momenta are as follows:
\begin{eqnarray}
\label{TB17} && P\ = p_1+p_2\ = p'_1+p'_2\ , \;\; k=\frac12(p_1-p_2),
\nn \\ && g^\perp_{\nu\mu}=g_{\nu\mu}-\frac{P_\nu P_\mu}{P^2}\equiv\ g^{\perp
P}_{\mu\nu}
\ , \;\;
k_\perp=k_\nu g^\perp_{\nu\mu}\ .
\end{eqnarray}

The representation (\ref{TB14}) is suitable to
consider  the $t$-channel meson or Reggeon exchanges, while
Eq. (\ref{TB16}) is convenient for the $s$-channel
partial-wave analysis. The re\-presentations (\ref{TB14}) and
(\ref{TB16}) are related to each other by the Fierz transformation
\cite{fierz}, with a corresponding re-expansion of the spin--momentum
operators.

In terms of the  $SLJ$ representations, the states are usually described
as $^{2S+1}L_J$. The $P$-parity can be calculated as $P=(-1)^{L+1}$
and $C=(-1)^{L+S}$. The states with $S=0$ are unambiguously
defined and they form a set of states with
$J^{PC}=0^{-+},1^{+-},2^{-+}\ldots$ The states with $S=1$ and $L=J$
are also uniquely defined and form the set
$J^{PC}=1^{++},2^{--},3^{++}\ldots$ The states with $S=1$ and
$L=J-1$ and $L=J+1$ have the same $J^{PC}$ and can mix with each other,
that are the states $J^{PC}=0^{++},1^{--},2^{++},\ldots$

\subsection{Nucleon-antinucleon scattering amplitude}

Let us write the $s$-channel expansion for a pair of nucleons
where $N=(p,n)$ forms an isodoublet. The systems  $p\bar n$ and $n\bar
p$ have isospin $I=1$, and the $s$-channel expansions of their
scattering amplitudes are determined by formulae which are analogous to
those for $\Lambda\bar\Lambda$, Eq. (\ref{TB16}). The systems  $p\bar
p$ and $n\bar n$ are a superposition of two states, with $I=0$ and
$I=1$. The nucleon--antinucleon amplitudes read:
\begin{eqnarray}
\label{TB22a}
&&\mbox{\boldmath$\!\!\!\!\!\!
p(p_1)\bar n(p_2)\to p(p'_1)\bar n(p'_2)\ (I=1):$}
\nonumber \\
&&\left(C^{11}_{1/2\ 1/2\ , \ 1/2\ 1/2}\right)^2
M_{1}(s,t,u) = M_{1}(s,t,u)\ ,
\\
\label{TB22b}
&&\mbox{\boldmath$\!\!\!\!\!\!
p(p_1)\bar p(p_2)\to p(p'_1)\bar p(p'_2)\ (I=0,1):$}
\nonumber\\
 &&\left(C^{10}_{1/2\ 1/2\ , \ 1/2\ -1/2}\right)^2  M_{1}(s,t,u)
\nonumber\\
&&+\left(C^{00}_{1/2\ 1/2\ , \ 1/2\ -1/2}\right)^2  M_{0}(s,t,u) =
\nonumber\\
&&\qquad\frac12 M_{1}(s,t,u) + \frac12 M_{0}(s,t,u)\ ,
\\
\label{TB22c}
&&\mbox{\boldmath$\!\!\!\!\!\!
p(p_1)\bar p(p_2)\to n(p'_1)\bar n(p'_2)\ (I=0,1):$}
\nonumber\\
 && C^{10}_{1/2\ 1/2\ , \ 1/2\ -1/2}\;C^{10}_{1/2\ -1/2\ , \ 1/2\ 1/2} M_{1}(s,t,u)
\nonumber\\
&&+ C^{00}_{1/2\ 1/2\ , \ 1/2\ -1/2}\;C^{00}_{1/2\ -1/2\ , \ 1/2\ 1/2} M_{0}(s,t,u)=
\nonumber\\
&&\qquad\frac12 M_{1}(s,t,u) - \frac12 M_{0}(s,t,u) \ .
\end{eqnarray}
Note that, by writing the $N\bar N$ (or $NN$) scattering amplitudes, one
can use alternatively either isotopic Pauli matrices ($I/\sqrt2$,
$\bta/\sqrt2$) or Clebsch-Gordan coefficients. In (\ref{TB22a}), we use
Clebsch-Gordan coefficients which allows us to consider reactions in
which states with $I>1/2$ are produced.

The $s$-channel operator expansion for $N\bar N\to N \bar N$ can be written as
\begin{eqnarray}
\label{TB23}
&& M_{I}(s,t,u) = \sum_{S,L,L',J \atop \mu_1\ldots\mu_J}
\left(\bar\psi(p'_1) Q^{SL'J}_{\mu_1\ldots\mu_J}(k')\psi^c(-p'_2)\right)
\nonumber\\
&&\times\left(\bar\psi^c(-p_2)Q^{SLJ}_{\mu_1\ldots\mu_J}(k)\psi(p_1)\right)
A^{(S,L'L,J)}_I(s) .
\end{eqnarray}
Since the two masses are equal, $k=k_\perp$ holds. In Eq. (\ref{TB23}),
the summation is performed over all states (as well as for the
$\Lambda\bar\Lambda$ scattering amplitude). The spin--momentum operators
$Q^{SLJ}_{\mu_1\ldots\mu_J}(k)$ for the states with $J=0,1,2$ are given
in Section 4.

\subsection{Amplitude for \boldmath$p\Lambda\to p\Lambda$\unboldmath\
scattering}

It is convenient to present  the amplitude $p\Lambda\to p\Lambda$
precisely in the same technique which  was used in the consideration of
the $s$-channel fermion--antifermion system. To this aim, we declare
$p$ being a fermion and $\Lambda $ an antifermion. Then, the
$s$-channel expansion for the $p\Lambda\to p\Lambda$ scattering
amplitude reads:
 \be
&&\!\!\!M_{N\Lambda\to N\Lambda}(s,t,u) =\!\!\!
\sum_{S,L,L',J \atop \mu_1\ldots\mu_J}\!
\left(\bar\psi_N (p'_1)
Q^{SL'J}_{\mu_1\ldots\mu_J}(k'_\perp)\psi^c_\Lambda(-p'_2)\right)
\nn \\
&&\times\left(\bar\psi^c_\Lambda(-p_2)
Q^{SLJ}_{\mu_1\ldots\mu_J}(k_\perp)\psi_N(p_1)\right)
A^{(S,L'L,J)}_{N\Lambda\to N\Lambda}(s)\ .
\label{TB25}
\ee

\subsection{Amplitude for \boldmath$\Lambda\Lambda\to
\Lambda\Lambda$\unboldmath\ scattering}

Let us present  the amplitude $\Lambda\Lambda\to \Lambda\Lambda$ in the
technique which was used  for reaction $p\Lambda\to p\Lambda $. So, we declare
the 1st $\Lambda $ to be a fermion and the 2nd one to be an antifermion. One
can distinguish between them, for example, in the c.m. system labeling  a
particle scattered into the backward hemisphere as "antifermion". Then the
$s$-channel expansion for the $\Lambda\Lambda\to\Lambda\Lambda$ scattering
amplitude reads:
\begin{eqnarray}
&&M_{\Lambda\Lambda\to \Lambda\Lambda}(s,t,u)=\!\!
\sum_{S,L,L',J \atop \mu_1\ldots\mu_J}\left(\bar\psi_\Lambda (p'_1)
Q^{SL'J}_{\mu_1\ldots\mu_J}(k')\psi^c_\Lambda(-p'_2)\right)
\nn \\
&&\times\left(\bar\psi^c_\Lambda(-p_2)
Q^{SLJ}_{\mu_1\ldots\mu_J}(k)\psi_\Lambda(p_1)\right)
A^{(S,L'L,J)}_{\Lambda\Lambda\to \Lambda\Lambda}(s)\ .
\label{TB36}
\end{eqnarray}
In this reaction, a selection rule for quantum numbers caused by the
Fermi statistics should be taken into account, such as:
\be
(-1)^{S+L+1}=-1.
\ee
Therefore, the following states contribute into (\ref{TB36}) only:
\be \label{TB37}
S=1&:&\; (L=1;J=1), \ (L=3;J=2,3,4),\ ...
\nonumber\\
S=0&:&\; (L=0;J=0), \ (L=2;J=2),\ ...
\ee

\subsection{Nucleon--nucleon scattering amplitude}

Nucleon is an isodoublet with components $p\to(I=1/2, I_3=1/2)$ and
$n\to(I=1/2,I_3=-1/2)$. The systems  $p p$ and $n n$ have total
isospin $I=1$, and the $s$-channel expansions of their scattering
amplitudes are determined by formulae analogous to those for
$\Lambda\Lambda$, Eq. (\ref{TB36}). The system  $p n$ is a  superposition
of two states, with total isospins $I=0$ and $I=1$. The amplitudes
read:
\begin{eqnarray}
&&\mbox{\boldmath$\!\!\!\!\!\!
p(p_1) p(p_2)\to p(p'_1) p(p'_2)\ (I=1): $}
\nonumber \\
&&\left(C^{11}_{1/2\ 1/2\ , \ 1/2\ 1/2}\right)^2
M_{1}(s,t,u)=  M_{1}(s,t,u)\ ,
\\
&&\mbox{\boldmath$\!\!\!\!\!\!
p(p_1) n(p_2)\to p(p'_1) n(p'_2)\ (I=0,1):$}
\nonumber \\
&& \left(C^{10}_{1/2\ 1/2\ , \ 1/2\ -1/2}\right)^2  M_{1}(s,t,u)
\nonumber\\
&&+\left(C^{00}_{1/2\ 1/2\ , \ 1/2\ -1/2}\right)^2  M_{0}(s,t,u) =
\nonumber \\
&&\qquad\frac12 M_{1}(s,t,u) + \frac12 M_{0}(s,t,u)\ ,
\\
&&\mbox{\boldmath$\!\!\!\!\!\!
n(p_1) n(p_2)\to n(p'_1) n(p'_2)\ (I=1):$}
\nonumber \\
&&\!\!\!\!\!\!
\left(C^{1-1}_{1/2\ -1/2\ , \ 1/2\ -1/2}\right)^2
M_{1}(s,t,u) =  M_{1}(s,t,u)\ .
\label{TB38}
\end{eqnarray}
The $s$-channel operator expansion gives for $M_{I}(s,t,u)$ in
the reaction $p n\to p n$ ($I=0$):
\begin{eqnarray}
&&\hspace{-0.5cm} M_{0}(s,t,u)=
\sum_{S,L,L',J \atop \mu_1\ldots\mu_J}
\left(\bar\psi_p
(p'_1) Q^{SL'J}_{\mu_1\ldots\mu_J}(k')\psi^c_n(-p'_2)\right)
\nonumber \\&&
\times \left(\bar\psi^c_n(-p_2)
Q^{SLJ}_{\mu_1\ldots\mu_J}(k)\psi_p(p_1)\right)
A^{(S,L'L,J)}_0(s) ,
\nn \\
S=1&:&\; (L=0;J=1), \ (L=2;J=1,2,3),\ ... \nonumber\\
S=0&:&\; (L=1;J=1), \ (L=3;J=3),\ ...
\label{TB39}
\end{eqnarray}
and for $I=1$:
\begin{eqnarray}
\label{TB40}
&&\hspace{-0.5cm} M_{1}(s,t,u)=
\sum_{S,L,L',J \atop \mu_1\ldots\mu_J}
\left(\bar\psi_p
(p'_1) Q^{SL'J}_{\mu_1\ldots\mu_J}(k')\psi^c_n(-p'_2)\right) \nonumber \\&&
\times
 \left(\bar\psi^c_n(-p_2)
 Q^{SLJ}_{\mu_1\ldots\mu_J}(k)\psi(p_1)\right)
A^{(S,L'L,J)}_1(s) , \nn \\
S=1&:&\quad (L=1;J=0,1,2), \ (L=3;J=2,3,4),\ ... \nonumber \\
S=0&:&\quad (L=0;J=0), \ (L=2;J=2),\ ...
\end{eqnarray}
The selection rule for quantum numbers in (\ref{TB39}) and (\ref{TB40})
is caused by the Fermi statistics.

Analogous partial wave expansions can be written for the reactions
$pp\to pp$ and $nn\to nn$ ($I=1$), with an obvious replacing in
(\ref{TB40}): $n\to p$ for $pp\to pp$ and $p\to n$ for $nn\to nn$.
Here, as for $\Lambda\Lambda\to\Lambda \Lambda $, declaring one nucleon
 as a fermion and the second one as antifermion, one  distinguishes
between them in c.m. system labeling  a particle scattered into the
backward hemisphere as "antifermion".


\section{Unitarity conditions and \boldmath$K$\unboldmath-matrix
representations of baryon--antibaryon and
 baryon--baryon  scattering amplitudes}

Here, we write down the unitarity conditions and give the $K$-matrix
representations of the baryon-antibaryon and baryon-baryon  scattering
amplitudes suggesting that inelastic processes are switched off (for
example, because the energy is not large enough). Generalisation of the
$K$-matrix representations in case when inelastic channels are
switched on can be performed in a standard way.

\subsection{\boldmath$\Lambda\bar\Lambda$\unboldmath\ scattering}

In this subsection, we consider the unitarity condition for the
amplitude with $J=L$. The generalisation for $J=L\pm 1$ amplitude
is considered in the last subsection. For the amplitude
$\Lambda\bar\Lambda\to\Lambda\bar\Lambda $ of Eq. (\ref{TB16}), the
$s$-channel unitarity condition reads for $J=L$ (we re-define
$A^{(S,LL,J)}_s(s)\to
 A^{(S,LL,J)}_{\Lambda\bar\Lambda\to\Lambda\bar\Lambda}(s)$)
 as follows:
\be
&& \sum_{ \mu_1\ldots\mu_J} \left(\bar\psi (p'_1)
Q^{SLJ}_{\mu_1\ldots\mu_J}(k')\psi^c(-p'_2)\right)
\nn \\
&&\times \left(\bar\psi^c(-p_2)
Q^{SLJ}_{\mu_1\ldots\mu_J}(k)\psi(p_1)\right)
{\it Im}\, A^{(S,LL,J)}_{\Lambda\bar\Lambda\to \Lambda\bar\Lambda}(s)=
\nn \\
&& \int d\Phi_2(p''_1,p''_2)
\sum_{j,\ell} \sum_{
\mu_1\ldots\mu_J} \left(\bar\psi (p'_1)
Q^{SLJ}_{\mu_1\ldots\mu_J}(k')\psi^c(-p'_2)\right)
\nn \\
&&\times\left(\bar\psi_\ell^c(-p''_2)
Q^{SLJ}_{\mu_1\ldots\mu_J}(k'')\psi_j(p''_1)\right)
A^{(S,LL,J)}_{\Lambda\bar\Lambda\to \Lambda\bar\Lambda}(s)
\nn \\
&&\times \sum_{ \mu''_1\ldots\mu''_J}\Big [ \left(\bar\psi (p_1)
Q^{SLJ}_{\mu''_1\ldots\mu''_J}(k)\psi^c(-p_2)\right)
\nn \\
&&\times \left(\bar\psi_\ell^c(-p''_2)
Q^{SLJ}_{\mu''_1\ldots\mu''_J}(k'')\psi_j(p''_1)
A^{(S,LL,J)}_{\Lambda\bar\Lambda\to \Lambda\bar\Lambda}(s)\right)
\Big ]^+  \ .
\label{TB41}
\ee
Finally, one has:
 \be
\!\!\!\!\!\!\!Im\, A^{(S,LL,J)}_{\Lambda\bar\Lambda\to \Lambda\bar\Lambda}(s)
\!=\!
\rho^{(SLJ)}_{\Lambda\bar\Lambda}(s)
A^{(S,LL,J)^*}_{\Lambda\bar\Lambda\to \Lambda\bar\Lambda}(s)
A^{(S,LL,J)}_{\Lambda\bar\Lambda\to \Lambda\bar\Lambda}(s),
\label{TB42}
\ee
where
\be
\label{TB43}
&&O^{\mu_1\ldots\mu_J}_{\mu''_1\ldots\mu''_J}
\rho^{(SLJ)}_{\Lambda\bar\Lambda}(s) =\int d\Phi_2(p''_1,p''_2)
\\
&& \times {\rm Sp}\left( Q^{SLJ}_{\mu_1\ldots\mu_J}(k'')
(-\hat p''_2+m_{\Lambda})   Q^{SLJ}_{\mu''_1\ldots\mu''_J}(k'')
(\hat p''_1+m_{\Lambda})\right)\, . \nonumber
\ee
The projection operator $O^{\mu_1\ldots\mu_J}_{\mu''_1\ldots\mu''_J}$
is presented in Section~4.
The phase space is determined as
\be
\label{TB44}
&&\!\!\!\!\!\!\!\!\!\!d\tilde\Phi_2(p_1,p_2)=  \frac{2\pi^4}{2}d\Phi_2(p_1,p_2) = \nn \\
&&\frac 12 (2\pi)^4 \delta^{(4)}(P-p_1-p_2)
\frac{d^3p_1}{(2\pi)^3 2p_{10}}\frac{d^3p_2}{(2\pi)^3 2p_{20}}.
\ee
The projection operator
$O^{\mu_1\ldots\mu_J}_{\mu''_1\ldots\mu''_J}$ obeys the convolution
rule, $O^{\mu_1\ldots\mu_J}_{\mu_1\ldots\mu_J}=2J+1$ , that gives:
\be
\label{TB46}
&&\rho^{(SLJ)}_{\Lambda\bar\Lambda}(s) =\frac{1}{2J+1} \int
d\tilde\Phi_2(p''_1,p''_2)
\\
&&\times  {\rm Sp}\Big(
Q^{SLJ}_{\mu_1\ldots\mu_J}(k'')(-\hat p''_2+m_{\Lambda})
Q^{SLJ}_{\mu_1\ldots\mu_J}(k'')(\hat p''_1+m_{\Lambda})\Big)\, .
\nonumber
\ee
The unitarity condition (\ref{TB42}) results in the following
$K$-matrix representation of the amplitude
$\Lambda\bar\Lambda\to \Lambda\bar\Lambda$:
 \be
\label{TB47}
  A^{(S,LL,J)}_{\Lambda\bar\Lambda\to \Lambda\bar\Lambda}(s)=
\frac {K^{(S,LL,J)}_{\Lambda\bar\Lambda\to \Lambda\bar\Lambda}(s)}{1-
i\rho^{(SLJ)}_{\Lambda\bar\Lambda}(s)
K^{(S,LL,J)}_{\Lambda\bar\Lambda\to \Lambda\bar\Lambda}(s)}.
\ee

\subsection{\boldmath$\Lambda\Lambda$\unboldmath\ scattering}

Likewise, we consider the unitarity condition for the $\Lambda\Lambda$
scattering amplitude. The $s$-channel unitarity condition for the
amplitude $\Lambda\Lambda\to \Lambda\Lambda$ with $J=L$ reads
 \be
\label{TB53}
 {\it Im}\, A^{(S,LL,J)}_{\Lambda\Lambda\to \Lambda\Lambda}(s)= \frac
 12
\rho^{(SLJ)}_{\Lambda\Lambda}(s)
A^{(S,LL,J)^*}_{\Lambda\Lambda\to \Lambda\Lambda}(s)
 A^{(S,LL,J)}_{\Lambda\Lambda\to \Lambda\Lambda}(s)\, ,\nn
\ee
where the identity factor $1/2$ is introduced. In this way, we
 keep the definition (\ref{TB44}) for $d\tilde\Phi_2(p''_1,p''_2)$.

We have
\be \label{TB54}
&&O^{\mu_1\ldots\mu_J}_{\mu''_1\ldots\mu''_J}
\rho^{(SLJ)}_{\Lambda\Lambda}(s) =\int d\tilde\Phi_2(p''_1,p''_2)
\\
&& \times {\rm Sp}\left(  Q^{SLJ}_{\mu_1\ldots\mu_J}(k'')(-\hat
p''_2+m_{\Lambda})   Q^{SLJ}_{\mu''_1\ldots\mu''_J}(k'')(\hat
p''_1+m_{\Lambda})\right)\, . \nonumber
\ee
The convolution rule
$O^{\mu_1\ldots\mu_J}_{\mu_1\ldots\mu_J}=2J+1$,
gives us
\be
\label{TB55}
&&\rho^{(SLJ)}_{\Lambda\Lambda}(s) =\frac{1}{2J+1} \int
d\tilde\Phi_2(p''_1,p''_2)
\\
&&\times  {\rm Sp}\left(
Q^{SLJ}_{\mu_1\ldots\mu_J}(k'')(-\hat p''_2+m_{\Lambda})
Q^{SLJ}_{\mu_1\ldots\mu_J}(k'')(\hat p''_1+m_{\Lambda})\right)\, ,
\nonumber
\ee
thus leading to identical definitions for
$\rho^{(SLJ)}_{\Lambda\Lambda}(s)\;$
and $\rho^{(SLJ)}_{\Lambda\bar\Lambda}(s)$, see (\ref{TB46}).
The unitarity condition (\ref{TB53}) results in the following
$K$-matrix representation of the amplitude
$\Lambda\Lambda\to \Lambda\Lambda$:
 \be
\label{TB56}
  A^{(S,LL,J)}_{\Lambda\Lambda\to \Lambda\Lambda}(s)=
\frac {K^{(S,LL,J)}_{\Lambda\Lambda\to \Lambda\Lambda}(s)}{1-
\frac i2 \rho^{(SLJ)}_{\Lambda\Lambda}(s)
K^{(S,LL,J)}_{\Lambda\Lambda\to \Lambda\Lambda}(s)}\ .
\ee
Let us emphasize the appearance of the identity factor $1/2$ in
the denominator of (\ref{TB56}).

\subsection{Nucleon--antinucleon partial wave amplitude}

The $K$-matrix representation for $N\bar N$
scattering amplitude is written precisely in the same way as for the
$\Lambda\bar\Lambda$ case. The only new aspect as compared to
$\Lambda\bar\Lambda$ is that the $N\bar N$ scattering is determined by
two isotopic amplitudes, see (\ref{TB22b}) and (\ref{TB22c}), with
$I=0,1$:
\be
\!\!\!\!\!\!\mbox{\boldmath$p\bar n\to p\bar n\ (I=1)$}\quad&:&
\frac12 M_{1}(s,t,u) +\frac12 M_{0}(s,t,u)\ ,
\nn \\
\!\!\!\!\!\!\mbox{\boldmath$p\bar p\to n\bar n\ (I=0,1)$}&:&
\frac12 M_{1}(s,t,u) -\frac12 M_{0}(s,t,u) .
\label{TB61}
\ee
Being expanded
over the $s$-channel operators
$Q^{SLJ}_{\mu_1\ldots\mu_J}(k) \otimes Q^{SL'J}_{\mu_1\ldots\mu_J}(k')$,
these amplitudes are represented $\,$
through partial wave amplitudes  $A^{(S,L'L,J)}_0(s)$ and
$A^{(S,L'L,J)}_1(s)$. The unitarity condition for these amplitudes
leads again to a $K$-matrix representation.

As above, we consider here the one-channel amplitude, first for $J=L$.
The two-channel amplitudes $(S=1,J=L\pm 1)$ are presented below in
Section 3.4.

The imaginary part of the amplitude $ A^{(S,LL,J)}_{I}(s)$ with $I=0,1$
and $J=L$ satisfying the $s$-channel unitarity condition reads
 \be
Im\, A^{(S,LL,J)}_{I}(s)\!=\!
\rho^{(S,LL,J)}_{N\bar N}(s)
A^{(S,LL,J)^*}_{I}(s)
A^{(S,LL,J)}_{I}(s), \nn \\
\label{TB62}
\ee
where
\be
\label{TB63}
&&\rho^{(S,LL',J)}_{N\bar N}(s) =\frac{1}{2J+1} \int
d\tilde\Phi_2(p_1,p_2)
\\
&& \times {\rm Sp}\left(
Q^{SLJ}_{\mu_1\ldots\mu_J}(k)(-\hat p_2+m_{N})
Q^{SL'J}_{\mu_1\ldots\mu_J}(k)(\hat p_1+m_{N})\right)\, . \nonumber
\ee
The unitarity condition (\ref{TB63}) gives us the following
$K$-matrix representation:
 \be
\label{TB64}
  A^{(S,LL,J)}_{I}(s)=
\frac {K^{(S,LL,J)}_{I}(s)}{1-
 i\ \rho^{(S,LL,J)}_{N\bar N}(s)
K^{(S,LL,J)}_{I}(s)}  \ .
\ee

\subsection{Nucleon--nucleon scattering amplitude}

 The  $p p$ and $n n$ systems are pure $I=1$ states, while the $pn$
is a  superposition of two states with total isospins $I=0$ and
$I=1$. The amplitudes read:
\begin{eqnarray}
\label{TB68}
\mbox{\boldmath$p p\!\to\! p p, \ nn\!\to\! nn\ (I\!\!=\!1)$}&:&
M_{1}(s,t,u) , \\
\mbox{\boldmath$p n\!\to\! p n \ (I\!\!=\!0,1)$}&:&
\frac12 M_{1}(s,t,u)\! +\! \frac12 M_{0}(s,t,u).\nn
\end{eqnarray}

The expansion over  $s$-channel operators
$Q^{SLJ}_{\mu_1\ldots\mu_J}(k)\otimes$ $Q^{SL'J}_{\mu_1\ldots\mu_J}\!(k')$
is a representation of these
amplitudes  through partial wave amplitudes  $A^{(S,L'L,J)}_0(s)$
and $A^{(S,L'L,J)}_1(s)$.


\noindent{\bf (i) Partial wave amplitudes $N N \to N N $ for $J=L$.} \\
For $J=L$, the amplitude $ A^{(S,LL,J)}_{I}(s)$ with $I=0,1$ sa\-tisfying
the $s$-chan\-nel unitarity condition are identical to those
for nucleon-antinucleon scattering, eqs. (\ref{TB62}) and
(\ref{TB63}), except for the factor $\frac{1}{2}$ in the amplitude
originating from Fermi-Dirac statistics.
 \be
\label{TB64_1}
  A^{(S,LL,J)}_{I}(s)=
\frac {K^{(S,LL,J)}_{I}(s)}{1- \frac i2 \ \rho^{(S,LL,J)}_{N N}(s)
K^{(S,LL,J)}_{I}(s)}  \ .
\ee

\noindent{\bf (ii) Partial wave amplitudes for $S=1, J=L\pm 1$.} \\
In this case, four partial amplitudes
  form a $2\times 2$ matrix given by
\be \label{TB65}
&&\hspace{-8mm}\widehat A^{(S=1,L=J\pm 1,J)}_{I}(s) \ = \nn \\ &&
\left |\begin{array}{cc}
 A^{(S=1,J- 1\to J-1,J)}_{I}(s),
 &
 A^{(S=1,J- 1\to J+1,J)}_{I}(s)
 \\
 A^{(S=1,J+ 1\to J-1,J)}_{I}(s),
  &
 A^{(S=1,J+ 1\to J+1,J)}_{I}(s)
 \end{array} \right |.
\ee
The $K$-matrix representation reads:
 \be
\label{TB66}
&&\hspace{-6mm}\widehat  A^{(S=1,L=J\pm 1,J)}_{I}(s)=
\widehat K^{(S=1,L=J\pm 1,J)}_{I}(s) \nn \\ && \times
\left [I-\frac i2\
\widehat\rho^{(S=1,L=J\pm 1,J)}_{N N}(s) \widehat
K^{(S=1,L=J\pm 1,J)}_{I}(s)\right ]^{-1} \,
\ee
with the following definitions:
\be \label{TB59}
&&\hspace{-8mm}\widehat K^{(S=1,L=J\pm 1,J)}_{I}(s) =  \nn \\ &&
 \left |\begin{array}{cc}
 K^{(S=1,J- 1\to J-1,J)}_{I}(s),
 &
K^{(S=1,J- 1\to J+1,J)}_{I}(s)
 \\
 K^{(S=1,J+ 1\to J-1,J)}_{I}(s),
  &
 K^{(S=1,J+ 1\to J+1,J)}_{I}(s)
 \end{array} \right |,\nn \\
&&\hspace{-8mm}\widehat \rho^{(S=1,L=J\pm 1,J)}_{N N}(s) =  \nn \\ &&
 \left |\begin{array}{cc}
 \rho^{(S=1,J- 1\to J-1,J)}_{N N}(s),
 &
\rho^{(S=1,J- 1\to J+1,J)}_{N N}(s)
 \\
 \rho^{(S=1,J+ 1\to J-1,J)}_{N N}(s),
  &
 \rho^{(S=1,J+ 1\to J+1,J)}_{N N}(s)
 \end{array} \right | .
\ee
Note  that the matrices
$\widehat \rho^{(S=1,L=J\pm 1,J)}_{I}(s) $ and \\
$\widehat K^{(S=1,L=J\pm 1,J)}_{I}(s)$ are symmetrical:
\be
&&
\rho^{(S=1,J- 1\to J+1,J)}_{N N}(s)=
\rho^{(S=1,J+ 1\to J-1,J)}_{N N}(s) \;,\nn \\ &&
K^{(S=1,J- 1\to J+1,J)}_{I}(s)=
K^{(S=1,J+ 1\to J-1,J)}_{I}(s).
\ee
Let us emphasize that the definitions of the phase spaces for
$NN$ and $N\bar N$ systems coincide:
$\rho^{(S,L\to L', J)}_{NN}(s)=$ \\
$\rho^{(S,L\to L', J)}_{N\bar N}(s)$.
In the equation imposing the unitarity condition (as well as in the
$K$-matrix representation), the identity of particles in the $NN$
systems is taken into account directly by the factor $1/2$. The
unitarity conditions for the $\Lambda \bar \Lambda$, $\Lambda \Lambda$
and $N \bar N$ two-channel partial wave amplitudes for $S=1$ and $J=L\pm
1$ are written similarly.

\section{Nucleon--nucleon interaction operators}

In this Section, the proton--proton interaction operators are
constructed. These operators are constructed using angular momentum and
spin operators, whose properties are discussed below.

\subsection{Angular momentum operators}

The angular-dependent part of the wave function of the composite state
is described by operators constructed  using relative momenta
of particles and the
metric tensor. Such operators (we  denote them as
$X^{(L)}_{\mu_1\ldots\mu_L}$, where $L$ is the angular momentum) are
called angular momentum operators; they correspond to irreducible
representations of the Lorentz group \cite{xoper,decompos}.
They satisfy the following properties \cite{xoper}:
(i) Symmetry with respect to permutation of any two indices:
\be
X^{(L)}_{\mu_1\ldots\mu_i\ldots\mu_j\ldots\mu_L}\; =\;
X^{(L)}_{\mu_1\ldots\mu_j\ldots\mu_i\ldots\mu_L}.
\label{oth_b1}
\ee
(ii) Orthogonality to the total momentum of the system, $P=k_1+k_2$:
\be
P_{\mu_i}X^{(L)}_{\mu_1\ldots\mu_i\ldots\mu_L}\ =\ 0.
\label{oth_b2}
\ee
The traceless property for the summation over two any indices:
\be
g_{\mu_i\mu_j}X^{(L)} _{\mu_1\ldots\mu_i\ldots\mu_j\ldots\mu_L}\
\ =\ 0.
\label{oth_b3}
\ee

Let us consider a one-loop diagram describing the decay of a composite
system into two spinless particles which propagate and then form again
a composite system. The decay and formation processes are described by
angular momentum operators. Due to the conservation of quantum
numbers, this amplitude must vanish for initial and final states with
different spin. The S-wave operator is a scalar and can be taken as a
unit operator. The P-wave operator is a vector. In the dispersion relation
approach, it is sufficient that the imaginary part of the loop
diagram, with S and P-wave operators as vertices, is equal to 0.
In the case of spinless particles this requirement entails
\be
\int\frac{d\Omega}{4\pi} X^{(1)}_\mu =0\ ,
\ee
where the integral is taken over the solid angle of the
relative momentum. In general, the result of such an integration
is proportional to the total momentum of the system $P_\mu$ (the only
external vector):
\be
\int\frac{d\Omega}{4\pi} X^{(1)}_\mu =\lambda P_\mu\;.
\ee
Convolution this expression with $P_\mu$ and demanding $\lambda=0$,
we obtain the orthogonality condition (\ref{oth_b2}).
The orthogonality between the D- and S-waves is provided by the
traceless condition (\ref{oth_b3}); conditions
(\ref{oth_b2}), (\ref{oth_b3}) provide the orthogonality for all
operators with different  angular momenta.

The orthogonality condition (\ref{oth_b2}) is automatically fulfilled
if the operators are constructed from the relative momenta
$k^\perp_\mu$ and tensor $g^\perp_{\mu\nu}$. Both of them are
orthogonal to the total momentum of the system, see eq.(\ref{TB17}).
In the c.m. system,
where $P=(P_0,\vec P)=(\sqrt s,0)$, the vector
$k^\perp$ is space-like: $k^\perp=(0,\vec k)$.

The operator for $L=0$ is a scalar (for example a unit
operator),
and the operator for $L=1$ is a vector, which can be
constructed from $k^\perp_\mu$ only.
The orbital angular momentum operators for $L =0 $ to 3
are:
\begin{eqnarray}
X^{(0)}&=&1\ , \qquad X^{(1)}_\mu=k^\perp_\mu\ , \qquad\nonumber \\
X^{(2)}_{\mu_1 \mu_2}&=&\frac32\left(k^\perp_{\mu_1}
k^\perp_{\mu_2}-\frac13\, k^2_\perp g^\perp_{\mu_1\mu_2}\right), \nonumber  \\
X^{(3)}_{\mu_1\mu_2\mu_3}&=& \\
\frac52\Big[k^\perp_{\mu_1} k^\perp_{\mu_2 } k^\perp_{\mu_3} &-&
\frac{k^2_\perp}5\left(g^\perp_{\mu_1\mu_2}k^\perp
_{\mu_3}+g^\perp_{\mu_1\mu_3}k^\perp_{\mu_2}+
g^\perp_{\mu_2\mu_3}k^\perp_{\mu_1} \right)\Big] . \nonumber \ee The operators
$X^{(L)}_{\mu_1\ldots\mu_L}$ for $L\ge 1$ can be written in the form of a
recurrent relation: \be X^{(L)}_{\mu_1\ldots\mu_L}&=&k^\perp_\alpha
Z^{\alpha}_{\mu_1\ldots\mu_L} \; ,
\nonumber\\
Z^{\alpha}_{\mu_1\ldots\mu_L}&=&
\frac{2L-1}{L^2}\Big (
\sum^L_{i=1}X^{{(L-1)}}_{\mu_1\ldots\mu_{i-1}\mu_{i+1}\ldots\mu_L}
g^\perp_{\mu_i\alpha}
\nonumber \\
 -\frac{2}{2L-1}  \sum^L_{i,j=1 \atop i<j}
&g^\perp_{\mu_i\mu_j}&
X^{{(L-1)}}_{\mu_1\ldots\mu_{i-1}\mu_{i+1}\ldots\mu_{j-1}\mu_{j+1}
\ldots\mu_L\alpha} \Big ).
\label{z}
\ee
The convolution equality reads
\be
X^{(L)}_{\mu_1\ldots\mu_L}k^\perp_{\mu_L}=k^2_\perp
X^{(L-1)}_{\mu_1\ldots\mu_{L-1}}.
\label{ceq}
\ee
Based on Eq.(\ref{ceq}) and
taking into account the traceless property of $X^{(L)}_{\mu_1\ldots\mu_L}$,
one can write down the orthogonali\-ty-normalisation condition for
orbital angular  operators
\begin{eqnarray}
&&\int\frac{d\Omega}{4\pi}
X^{(L)}_{\mu_1\ldots\mu_L}(k^\perp)X^{(L')}_{\mu_1\ldots\mu_{L'}}(k^\perp)
\ =\ \delta_{LL'}\alpha_L k^{2L}_\perp \; , \label{ort-x}\nn
\\
&&\alpha_L\ =\ \prod^L_{l=1}\frac{2l-1}{l} \; .
\label{alpha}
\ee
Iterating Eq. (\ref{z}), one obtains the
following expression for the operator $
X^{(L)}_{\mu_1\ldots\mu_L}$:
\be
\label{x-direct}
X^{(L)}_{\mu_1\ldots\mu_L}(k^\perp)=
\alpha_L &\bigg [&
k^\perp_{\mu_1}k^\perp_{\mu_2}k^\perp_{\mu_3}k^\perp_{\mu_4}
\ldots k^\perp_{\mu_L}
\nn \\
&-&\frac{k^2_\perp}{2L-1}\bigg(
g^\perp_{\mu_1\mu_2}k^\perp_{\mu_3}k^\perp_{\mu_4}\ldots
k^\perp_{\mu_L}
\nn \\
&&\qquad\;+g^\perp_{\mu_1\mu_3}k^\perp_{\mu_2}k^\perp_{\mu_4}\ldots
k^\perp_{\mu_L} + \ldots \bigg)
\nn \\
+\frac{k^4_\perp}{(2L\!-\!1)(2L\!-\!3)}&\bigg(&
g^\perp_{\mu_1\mu_2}g^\perp_{\mu_3\mu_4}k^\perp_{\mu_5}
k^\perp_{\mu_6}\ldots k^\perp_{\mu_L}
\nn \\
+
g^\perp_{\mu_1\mu_2}g^\perp_{\mu_3\mu_5}&k^\perp_{\mu_4}
&k^\perp_{\mu_6}\ldots k^\perp_{\mu_L}+
\ldots\bigg)+\ldots\bigg ].
\ee

\subsection{Projection operators and boson propagator}

The projection operator $O^{\mu_1\ldots\mu_L}_{\nu_1\ldots \nu_L}$
is constructed from the metric tensors $g^\perp_{\mu\nu}$ and it
has the following properties:
\be
X^{(L)}_{\mu_1\ldots\mu_L}
O^{\mu_1\ldots\mu_L}_{\nu_1\ldots \nu_L}\
&=&\ X^{(L)}_{\nu_1\ldots \nu_L}\ , \nonumber \\
O^{\mu_1\ldots\mu_L}_{\alpha_1\ldots\alpha_L} \
O^{\alpha_1\ldots\alpha_L}_{\nu_1\ldots \nu_L}\
&=& O^{\mu_1\ldots\mu_L}_{\nu_1\ldots \nu_L}\ .
\label{proj_op}
\ee
Taking into account the definition of the projection operators
(\ref{proj_op})
and the properties of the $X$-operators (\ref{x-direct}), we obtain
\be
k_{\mu_1}\ldots k_{\mu_L} O^{\mu_1\ldots\mu_L}_{\nu_1\ldots \nu_L}\
= \frac{1}{\alpha_L} X^{(L)}_{\nu_1\ldots\nu_L}(k^\perp).
\label{19}
\ee
This equation presents the basic property of projection operator:
it projects any operator with $L$ indices onto the partial
wave operator with angular momentum $L$.

For the lowest states,
\be
&&\hspace{-6mm}O= 1\ ,\qquad O^\mu_\nu=g_{\mu\nu}^\perp
\nn \\
&&\hspace{-6mm}O^{\mu_1\mu_2}_{\nu_1\nu_2}=
\frac 12 \left (
g_{\mu_1\nu_1}^\perp  g_{\mu_2\nu_2}^\perp \!+\!
g_{\mu_1\nu_2}^\perp  g_{\mu_2\nu_1}^\perp  \!- \!\frac 23
g_{\mu_1\mu_2}^\perp  g_{\nu_1\nu_2}^\perp \right ).\;\;
\ee
For higher states, the operator can be calculated using the
recurrent expression
\be O^{\mu_1\ldots\mu_L}_{\nu_1\ldots\nu_L}&=&
\frac{1}{L^2} \bigg (
\sum\limits_{i,j=1}^{L}g^\perp_{\mu_i\nu_j}
O^{\mu_1\ldots\mu_{i-1}\mu_{i+1}\ldots\mu_L}_{\nu_1\ldots
\nu_{j-1}\nu_{j+1}\ldots\nu_L}
 \\
&-& \frac{4}{(2L-1)(2L-3)}
\nn \\
&\times&
\sum\limits_{i<j\atop k<m}^{L}
g^\perp_{\mu_i\mu_j}g^\perp_{\nu_k\nu_m}
O^{\mu_1\ldots\mu_{i-1}\mu_{i+1}\ldots\mu_{j-1}\mu_{j+1}\ldots\mu_L}_
{\nu_1\ldots\nu_{k-1}\nu_{k+1}\ldots\nu_{m-1}\nu_{m+1}\ldots\nu_L}
\bigg ).\nn
\ee

The product of two $X$-operators integrated over solid
angle (which is equivalent to an integration over internal momenta)
depends on external momenta and the metric tensor only. Therefore, it
must be proportional to the projection operator.
After straightforward calculations, we obtain
\be
\!\!\!\int\!\frac{d\Omega }{4\pi}\!
X^{(L)}_{\mu_1\ldots\mu_L}(k^\perp)
X^{(L)}_{\nu_1\ldots\nu_L}(k^\perp)\!=\!
\frac{\alpha_L\, k^{2L}_\perp}{2L\!+\!1}
O^{\mu_1\ldots\mu_L}_{\nu_1\ldots \nu_L}\ .
\label{x-prod}
\ee
Let us introduce the positive value $|\vec k|^2$:
\be
|\vec k|^2\!=\!-k_\perp^2\!=\!
\frac{[s\!-\!(m_1\!+\!m_2)^2][s\!-\!(m_1\!-\!m_2)^2]}{4s}\ .
\label{k2_rel}
\ee
In the c.m.s. of the reaction, $\vec k$ is the momentum of a
particle. In other systems, we use this definition only in the sense
of $|\vec k|\equiv \sqrt{-k_\perp^2}$; clearly, $|\vec k|^2$ is a
relativistically invariant positive value. Then, Eq. (\ref{x-prod}) can
be written as
\be
\int\!\frac{d\Omega }{4\pi}\!
X^{(L)}_{\mu_1\ldots\mu_L}(k^\perp)
X^{(L)}_{\nu_1\ldots\nu_L}(k^\perp)\!=\!
\frac{\alpha_L\,|\vec k|^{2L}}{2L\!+\!1}(-1)^L
O^{\mu_1\ldots\mu_L}_{\nu_1\ldots \nu_L}. \nn \\
\ee

The tensor part of numerator of the boson propagator is defined by the projection
operator. Let us write it as
\be
F^{\mu_1\ldots\mu_L}_{\nu_1\ldots\nu_L}=
(-1)^L\,O^{\mu_1\ldots\mu_L}_{\nu_1\ldots \nu_L}\ .
\label{boson_prop}
\ee
This definition guarantees that the width of a resonance (calculated
using the decay vertices) has a positive value.

\subsection{Spin operators of two-fermion systems}

The wave function for fermion particles with the momentum $p$ is described as Dirac
bispinor:
 \be
u(p)&=&\frac{1}{\sqrt{2m}\sqrt{p_0+m}} \left (
 \begin{array}{c}
(p_0+m)\omega \\
(\vec p\vec \sigma)\omega
 \end{array}
\right ),
\nn \\
\bar u(p)&=&\frac{ \left ( \omega^* (p_0+m),
-\omega^*(\vec p\vec \sigma) \right ) }{\sqrt{2m}\sqrt{p_0+m}}.
 \ee

To construct the operators for the  two-fermion system, one should also introduce the
charge-conjugated bispinors:
 \be
u(-p)&=&\frac{i}{\sqrt{2m}\sqrt{p_0+m}} \left (
 \begin{array}{c}
(\vec p\vec \sigma)\omega' \\
(p_0+m)\omega'
 \end{array}
\right ),
\nn \\
\bar u(-p)&=&-i\frac{ \left ( \omega'^*(\vec p\vec
\sigma)\ , \ -\omega'^* (p_0+m), \right ) }{\sqrt{2m}\sqrt{p_0+m}}.
 \ee
Here, the $\omega$ and $\omega'$ represent  2-dimensional spinors,
$\omega^*$ and $\omega'^*$ are the conjugated and transposed spinors. The
normalisation condition can be written as
 \be \bar u(p) u(p)&=& -\bar u(-p) u(-p) = 1\ ,  \nonumber \\
\sum\limits_{polarisations} u(p) \bar u(p)&=&\frac{m+\hat p}{2m}\ , \nn \\
\sum\limits_{polarisations} u(-p) \bar u(-p)&=&\frac{-m+\hat p}{2m}\ ,
\label{bisp_norm}
 \ee
where $\hat p =p^\mu\gamma_\mu$.

Let us consider a two-fermion system with the total momentum
$P=k_1+k_2$ and relative momentum $k=(k_1-k_2)/2$, where $k_1$ and
$k_2$ are their individual momenta, $P^2=s$. For the sake of generality,
let the fermions have different masses, $m_1$ and $m_2$. The
two-fermion system can form two possible spin state, $S=0$ (singlet
state) and $S=1$ (triplet state). The  spin operators for these
states act between bispinor and charge-conjugated bispinor,\\ $\bar
u(-k_1) S^{(i)} u(k_2)$ and have the following form:
 \be
S^{(0)}&=&i\gamma_5\ ,\quad S^{(1)}=\gamma^\perp_\mu \ ,
\label{spin_oper}
\ee
where
\be
\gamma^\perp_\mu&=&\gamma_\nu \;g_{\mu\nu}^\perp .
\ee
It should be noted that $u(-k_1)$ and $u(k_2)$ have opposite parities, so
$\bar u(-k_1) \gamma_5 u(k_2)$ is a scalar and $\bar u(-k_1) u(k_2)$ is a pseudoscalar.

As is shown below, that the $\gamma_\mu$ operator leads to the
mixture of states with total momentum $L+1$ and $L-1$. So, let us
introduce  the operator for the pure $S=1$ state:
 \be
S_{pure}^{(1)} = \Gamma^\perp_\alpha=\gamma_\beta\left(g^\perp_{\alpha\beta}
-\frac{4s k^\perp_\alpha k^\perp_\beta}
{M(\sqrt{s}+M)(s-\delta^2)}\right) ,\label{purespinstate}
 \ee
where $M=m_1+m_2$ and $\delta=m_1-m_2$. In the nonrelativistic
limit, this operator is equal to the spin-1 operator $\vec\sigma$ and
satisfies the orthogonality of the triplet states with the same parity.

\subsection{Operators for {\boldmath $^1L_J$} states}

In case of a singlet spin state, the total angular momentum $J$ is
equal to the orbital angular momentum $L$ between the two particles.
The ground state of such a system is $^1S_0$ $(^{2S+1}L_J)$ and
corresponding operator is just equal to the spin-0 operator $S^{(0)}$ of
Eq.(\ref{spin_oper}). For states with orbital momentum $L$, the
operator is constructed as a product of the spin-0 operator $S^{(0)}$
and the angular momentum operator $X_{\mu_1\ldots\mu_J}$:
\be
V_{\mu_1\ldots\mu_J}=\sqrt{\frac{2J+1}{\alpha_J}} i\gamma_5
X^{(J)}_{\mu_1\ldots\mu_J}(k^\perp)\;.
\label{V_1Lj}
 \ee
The normalisation factor which is introduced here simplifies
the expression for the loop
diagram (see below).

\subsection{Operators for {\boldmath $^3L_J$} states with {\boldmath $J\!=\!L$}}

The ground state in this series is $^3P_1$, so one should make a
convolution of two vectors, $S^{(1)}_\mu$ and $X^{(1)}_\mu$ that
creates a $J=1$ state (vector state). In this case the vertex operator
is equal to  $\varepsilon_{\nu_1\eta\xi\gamma}\gamma_\eta k^\perp_\xi
P_\gamma$. For states with higher orbital momenta, one needs to replace
$k^\perp_\xi$ by  $X^{(J)}_{\xi\nu_2\ldots\nu_J}$  and perform a full
symmetrisation over $\nu_1,\nu_2,\ldots,\nu_J$ indices, which can be
done by a convolution with projection operator
$O^{\mu_1\ldots\mu_L}_{\nu_1\ldots\nu_L}$. The general form of such
a vertex is hence given by
 \be
V^{L=J}_{\mu_1\ldots\mu_J}
\sim
\varepsilon_{\nu_1\eta\xi\gamma}\gamma_\eta
X^{(J)}_{\xi\nu_2\ldots\nu_J}P_\gamma O^{\mu_1\ldots\mu_J}_{\nu_1\ldots\nu_J}\ .
\label{V_3Lj_1}
 \ee
Using eqs.(\ref{x-direct}) and (\ref{19}), one has
\be
\varepsilon_{\nu_1\eta\xi\gamma}
X^{(J)}_{\xi\nu_2\ldots\nu_J}O^{\mu_1\ldots\mu_J}_{\nu_1\ldots\nu_J}
&=&
\varepsilon_{\nu_1\eta\xi\gamma} k^\perp_\xi
X^{(J-1)}_{\nu_2\ldots\nu_J}O^{\mu_1\ldots\mu_J}_{\nu_1\ldots\nu_J}\nn \\
&\times& \frac{2J-1}{J}\ .
\ee
Finally, using Eq. (\ref{App_1}) the vertex operator can be written as:
\be
V^{L=J}_{\mu_1\ldots\mu_J}
= \sqrt{\frac{(2J+1)J}{(J+1)\alpha_J}}
\frac{i\varepsilon_{\alpha\eta\xi\gamma}\gamma_\eta
k^\perp_\xi P_\gamma Z^\alpha_{\mu_1\ldots\mu_J}}
{\sqrt s}\ ,
\label{V_3Lj_1_fin}
 \ee
where normalisation parameters are again introduced. No\-te
that due to the property of the antisymmetric tensor
$\varepsilon_{\alpha\eta\xi\gamma}$ the vertex given by Eq.
(\ref{V_3Lj_1_fin}) does not change, if one replaces $\gamma_\eta$ by
the pure spin operator $\Gamma_\eta$.

\subsection{Operators for {\boldmath $^3L_J$} states with {\boldmath $L\!<\!J$}
and {\boldmath $L\!>\!J$}}

To construct operators for $^3L_J$ states, one should multiply the
spin operator $\gamma_\alpha$ by the orbital momentum operator for $L=
J+1$. So one has:
 \be
V^{L<J}_{\mu_1\ldots\mu_J}&\sim&
\gamma_{\nu_1} X^{(J-1)}_{\nu_2\ldots\nu_J} O^{\mu_1\ldots\mu_J}_{\nu_1\ldots\nu_J}\ .
\ee
Using Eq.(\ref{App_1}) from Appendix 3, we write the vertex operator
in the form
 \be
V^{L<J}_{\mu_1\ldots\mu_J}&=&
\gamma_\alpha Z^{\alpha}_{\mu_1\ldots\mu_J}
\sqrt{\frac{J}{\alpha_J}}\ ,
\label{V_3Lj_2}
\ee
and for the pure spin operator as
 \be
\tilde V^{L<J}_{\mu_1\ldots\mu_J}&=&
\Gamma_\alpha Z^{\alpha}_{\mu_1\ldots\mu_J}
\sqrt{\frac{J}{\alpha_J}}\ .
\label{V_3Lj_2pure}
\ee
   The normalisation constant is chosen to facilitate the
calculation of loop diagrams containing such a vertex.

To construct such an operator for $L>J$
one should reduce the number of indices in the orbital operator by
a convolution with the spin operator:
 \be V^{L>J}_{\mu_1\ldots\mu_J}&=& \gamma_\alpha X_{\alpha\mu_1\ldots\mu_J}
\sqrt{\frac{J+1}{\alpha_J}}\ ,
\label{V_3Lj_3}
 \ee
and for pure spin state:
 \be \tilde V^{L>J}_{\mu_1\ldots\mu_J}&=& \Gamma_\alpha X_{\alpha\mu_1\ldots\mu_J}
\sqrt{\frac{J+1}{\alpha_J}}\ .
\label{V_3Lj_3pure}
 \ee


\section{Calculation of the \boldmath$NN\rightarrow NN$\unboldmath\
amplitude} \subsection{Structure of the amplitude}

Let us consider $N(q_1)N(q_2)\to N(k_1)N(k_2)$ transition amplitude with $q_1$,
$q_2$, $k_1$, $k_2$ being the nucleon momenta, and
$k=(k_1-k_2)/2$, $q=(q_1-q_2)/2$. In this section the structure of such amplitude
for different initial and final states is derived.

We start by considering $NN\to NN$ amplitudes for a singlet state.
For the $^1L_J$ state with $L=J$, the amplitude is given by
\be A_s&=&
\bar u(-q_2) V_{\mu_1\ldots\mu_J}(q)\ u(q_1) \nn \\ &\times&
F^{\mu_1\ldots\mu_J}_{\nu_1\ldots\nu_J}\ \bar u(k_1)
V_{\nu_1\ldots\nu_J}(k)\ u(-k_2)\ .
\ee
For sake of simplicity, we omit here and below the invariant
part of the amplitude, which will be considered later on. Using Eq.
(\ref{V_1Lj}), the amplitude reads
 \be
A_s&=&-\bar u(-q_2)\ \gamma_5 \ u(q_1)\
    \bar u(k_1)\ \gamma_5 \ u(-k_2)\  \nn \\
&\times&(|\vec{k}|\ |\vec{q}|)^n    (2J+1) P_J(z)\ ,
 \ee
where $z=(\vec{k}\vec{q})/(|\vec{k}||\vec{q}|)$ is the cosine of
scattering angle in c.m. system.

The transition amplitude for the triplet state $^3L_J$ with
$L<J$ is the following:
\be
A_t^{L<J}&=& \bar u(-q_2)
V^{L<J}_{\mu_1\ldots\mu_J}(q)\ u(q_1) \nn \\
&\times&\ F^{\mu_1\ldots\mu_J}_{\nu_1\ldots\nu_J}\ \bar u(k_1)
V^{L<J}_{\nu_1\ldots\nu_J}(k)\ u(-k_2)\ .
\label{amplitude_tr1}
\ee
Using Eq. (\ref{App_1}), this amplitude can be written in the form
 \be
A_t^{L<J}=\sum\limits_{i=1}^{5}\ f_i\ a_i\ (|\vec{k}| |\vec{q}|)^{J-1},
 \ee
where
 \be
f_1&=&\bar u(-q_2)\ \gamma_\mu \ u(q_1)\
      \bar u(k_1)\ \gamma_\mu \ u(-k_2)\ ,\nn \\
f_2&=&\bar u(-q_2)\ \hat{q}\ u(q_1)\
      \bar u(k_1)\ \hat{q}\ u(-k_2)\;,\nn \\
f_3&=&\bar u(-q_2)\ \hat{k}\ u(q_1)\
      \bar u(k_1)\ \hat{k}\ u(-k_2) \;,\nn \\
f_4&=&\bar u(-q_2)\ \hat{q}\ u(q_1)\
      \bar u(k_1)\ \hat{k}\ u(-k_2)\;,\nn \\
f_5&=&\bar u(-q_2)\ \hat{k}\ u(q_1)\
      \bar u(k_1)\ \hat{q}\ u(-k_2)\ .
\ee
Here,
\be
a_1&=&-\frac{P'_J(z)}{J}\ ,\quad
a_2 = -\frac{P''_{J-1}(z)}{J|\vec{q}|^2}\ ,\quad
a_3= -\frac{P''_{J-1}(z)}{J|\vec{k}|^2}\ ,\nn \\
a_4&=&\! \frac{1}{J|\vec{k}||\vec{q}|}\ (P''_{J-2}(z) - 2P'_{J-1}(z))\ ,\;
a_5 = \frac{P''_J(z)}{J|\vec{k}||\vec{q}|}\ .\quad
 \ee
Likewise, the transition amplitude for triplet state $^3L_J$ with
$L>J$ is as follows:
 \be
A_t^{L>J}=\sum\limits_{i=1}^{5}\ f_i\ a_i\ (|\vec{k}| |\vec{q}|)^{J+1},
 \ee
where
 \be
a_1&=&-\frac{P'_{J+1}(z)}{J+1}\ ,\quad
a_2 = a_3= \frac{P''_{J+1}(z)}{J+1}\ , \\
a_4&\!=\!&-\frac{1}{J+1}\ (P''_J(z)\!+\!(2J+1)P'_{J+1}(z))\ ,\;
a_5 = -\frac{P''_J(z)}{J+1} .\nn
 \ee
If the spin-1 operator is defined as $\gamma_\nu$,
there is a mixture between  two triplet amplitudes
with $L>J$ and $L<J$. The corresponding transition amplitudes are
given by
 \be
A^{mix}_t&=&
V^{L<J}_{\mu_1\ldots\mu_J}(q)\
(-1)^J\ O^{\mu_1\ldots\mu_J}_{\nu_1\ldots\nu_J}\
V^{L>J}_{\nu_1\ldots\nu_J}(k)=
\nn \\
&&\sum\limits_{i=1}^{5}\ f_i\ a_i\ |\vec{k}|^{J+1} |\vec{q}|^{J-1},
\nn \\
A^{mix}_t&=&
V^{L>J}_{\mu_1\ldots\mu_J}(q)\
(-1)^J\ O^{\mu_1\ldots\mu_J}_{\nu_1\ldots\nu_J}\
V^{L<J}_{\nu_1\ldots\nu_J}(k)=
\nn \\
&&\sum\limits_{i=1}^{5}\ f_i\ a_i\ |\vec{k}|^{J-1} |\vec{q}|^{J+1},
 \ee
where
 \be
a_1&=&    -\sqrt{\frac{J}{J+1}}\frac{P'_J(z)}{2J-1}\ ,\quad
a_2 =      \sqrt{\frac{J}{J+1}}\frac{P''_{J-1}(z)}{2J-1}\ ,\\
a_3&=&     \sqrt{\frac{J}{J+1}}\frac{P''_{J+1}(z)}{2J-1}\ ,\quad
a_4 = a_5=-\sqrt{\frac{J}{J+1}}\frac{P''_J(z)}{2J-1}\ .\nonumber
 \ee
If one uses the operators based on the pure spin-1 operator given by
eqs. (\ref{V_3Lj_2pure}) and (\ref{V_3Lj_3pure}), the functions
$f_1,f_2,\ldots f_5$ are substituted by the new functions $\tilde
f_1,\tilde f_2,\ldots \tilde f_5$ as follows: \be \tilde f_i \;=\; f_j
M_{ji}\ , \ee where the transition matrix $M_{ji}$ is equal to \be
\left (
\begin{array}{ccccc}
1       & 0           & 0 &0 & 0\\
-\kappa & \frac{\sqrt{s}}{M} & 0 &0 & -\kappa (k^\perp q^\perp)\\
-\kappa & 0           & \frac{\sqrt{s}}{M} &0 & -\kappa (k^\perp q^\perp)\\
\kappa^2(k^\perp q^\perp)       & -\frac{\sqrt{s}\kappa(k^\perp q^\perp)}{M}          &
-\frac{\sqrt{s}\kappa(k^\perp q^\perp)}{M} &\frac{s}{M^2}& \kappa^2 (k^\perp q^\perp)^2\\
 0       & 0           & 0           &0          & 1\end{array} \right ).
\nonumber \\
\ee
Then, the transition amplitudes for the $^3L_J$ triplet state
 with
$L<J$  and $L<J$ are
\be
&& A_t^{L<J}=\sum\limits_{i,j=1}^{5}\ f_j M_{ji} \ a_i\
(|\vec{k}| |\vec{q}|)^{J-1},\nn \\
&& A_t^{L>J}=\sum\limits_{i,j=1}^{5}\ f_j M_{ji} \ a_i\ (|\vec{k}| |\vec{q}|)^{J+1}\ .
 \ee

The transition amplitude for the  $^3L_J$ triplet state with
$L=J$ is given by
\be
A_t^{L=J}&=& \bar u(-q_2)
V^{L=J}_{\mu_1\ldots\mu_J}(q)\ u(q_1)\
\nn \\
&\times& F^{\mu_1\ldots\mu_J}_{\nu_1\ldots\nu_J}\ \bar u(k_1)
V^{L=J}_{\nu_1\ldots\nu_J}(k)\ u(-k_2)\ .
\label{amplitude_tr1_1}
\ee
Using expressions given in Appendix 3, this amplitude can be written in
the form
 \be
A_t^{L=J}= (f_1 a_1 + f_5 a_5 +f_6 a_6)(|\vec{k}| |\vec{q}|)^{J},
 \ee
where
 \be
f_3&=&\bar u(-q_2)\ \gamma_\mu \ u(q_1)\
      \bar u(k_1)\ \gamma_\nu \ u(-k_2)\ n_\mu\ n_\nu, \\
n_\mu&=&\frac{\varepsilon_{\mu\alpha\beta\gamma}k_\alpha q_\beta P_\gamma}
             {\sqrt{s}\ |\vec{k}|\ |\vec{q}|}\ ,\quad \nn
a_1 = -\frac{2J+1}{(J+1)J}\ zP'_J(z)\ ,\\
a_5&=&-\frac{2J+1}{(J+1)J |\vec{k}|\ |\vec{q}|}\  P'_J(z)\ ,\quad\nn
a_6 =  \frac{2J+1}{(J+1)J}\  P''_J(z)\ .\nonumber
 \ee

\subsection{One-loop diagrams}

The calculation of one-loop diagrams for different vertex operators is
an important step in the construction of a unitary $NN$ amplitude. Let
us start from the loop diagram for the singlet state and derive all
expressions for the case of different particle masses, $m_1$ and $m_2$.
Taking into account that
 \be
Sp\left[\gamma_5(m_1+ \hat k_1)\gamma_5(m_2-\hat k_2)\right]=
2(s-\delta^2)\ ,
\label{gg_1}
 \ee
where $\delta = m_1 -m_2$, the one-loop diagram for the singlet state
is equal to
 \be
&&\!-\!\int\frac{d\Omega}{4\pi}Sp\Big[V_{\mu_1\ldots\mu_J}(k^\perp)(m_1+\hat k_1)
V_{\nu_1\ldots\nu_J}(k^\perp)(m_2-\hat k_2)\Big]
\nn \\
&&=2(s-\delta^2)|\vec k|^{2J}
O^{\mu_1\ldots\mu_J}_{\nu_1\ldots\nu_J}(-1)^J. \label{loop_singlet}
 \ee
The factor $(-1)$ is related to the fermionic nature of the baryon
in the loop.

To calculate one-loop diagrams for different triplet sta\-tes,
the following  relations are helpful:
 \be
Sp\Big[\gamma^\perp_\mu(m_1\!+\!\hat k_1)\gamma^\perp_\nu(m_2\!-\!\hat k_2)\Big]
&\!=&\! 2(s\!-\!\delta^2)g^\perp_{\mu\nu}\! + 8k^\perp_\mu k^\perp_\nu  ,
\nn \\
Sp\Big[\Gamma_\mu(m_1\!+\!\hat k_1)\Gamma_\nu(m_2\!-\!\hat k_2)\Big]
&\!=&\! 2(s\!-\!\delta^2)g^\perp_{\mu\nu}\ .
\label{gg_2}
 \ee
Using these relations and  eqs. (\ref{app_loop1})-(\ref{app_loop6})
 given in Appendix 3,
we obtain  the following results for the $L<J$ and $L>J$ states:
 \be
&&\hspace{-6mm}
-\int \frac{d\Omega}{4\pi} Sp\Big[V^{L<J}_{\mu_1\ldots\mu_J}(m_1+\hat
k_1) V^{L<J}_{\nu_1\ldots\nu_J}(m_2-\hat k_2)\Big]=
\nn \\
&&\hspace{-3mm}\left( 2(s\!-\!\delta^2)
\!-\!\frac{8J|\vec k|^2}{2J\!+\!1}\right)|\vec k|^{2(J-1)}
O^{\mu_1\ldots\mu_J}_{\nu_1\ldots\nu_J}(-1)^J \;,
\nn \\
&&\hspace{-6mm}
-\int \frac{d\Omega}{4\pi} Sp\Big[V^{L>J}_{\mu_1\ldots\mu_J}(m_1+\hat k_1)
V^{L>J}_{\nu_1\ldots\nu_J}(m_2-\hat k_2)\Big]=
\nn \\
&&\hspace{-3mm}\left( 2(s\!-\!\delta^2)
\!-\!\frac{8(J\!+\!1)|\vec k|^2}{2J\!+\!1}\right)|\vec k|^{2(J+1)}
O^{\mu_1\ldots\mu_J}_{\nu_1\ldots\nu_J}(-1)^J  .\quad
\label{loop_trip_1}
\ee
In case of spin-1 operators, the two triplet states with the same parity
are not orthogonal to each other; the interference loop diagram is
equal to
\be &&-\int \frac{d\Omega}{4\pi}
Sp\Big[V^{L<J}_{\mu_1\ldots\mu_J}(m_1+\hat k_1)
V^{L>J}_{\nu_1\ldots\nu_J}(m_2-\hat k_2)\Big]=
\nn \\
&&\qquad8\frac{\sqrt{J(J\!+\!1)}}{2J\!+\!1}|\vec k|^{2(J+1)}
O^{\mu_1\ldots\mu_J}_{\nu_1\ldots\nu_J}(-1)^J \ .
\label{loop_trip_2}
 \ee
The one-loop diagram for the $L=J$ triplet state is given by
 \be
&&\!-\!\int\frac{d\Omega}{4\pi}Sp\Big[V_{\mu_1\ldots\mu_J}(m_1+\hat
k_1)(k^\perp) V_{\nu_1\ldots\nu_J}(k^\perp)(m_2-\hat k_2)\Big] \nn \\
&&\!=2(s-\delta^2)|\vec k|^{2J}
O^{\mu_1\ldots\mu_J}_{\nu_1\ldots\nu_J}(-1)^J \ .
\label{loop_trip_3}
 \ee
Direct calculations show that the transition loop diagrams between the
triplet state with $L=J$ and the triplet states with $L>J$ and $L<J$
vanish. For vertex operators describing pure spin states
(\ref{V_3Lj_2pure}) and (\ref{V_3Lj_3pure}), one has the following
one-loop diagrams:
 \be
&&-\int \frac{d\Omega}{4\pi} Sp\Big[\tilde V^{L<J}_{\mu_1\ldots\mu_J}(m_1+\hat k_1)
\tilde V^{L<J}_{\nu_1\ldots\nu_J}(m_2-\hat k_2)\Big]=
\nn \\
&&\qquad 2(s\!-\!\delta^2)|\vec k|^{2(J-1)}
O^{\mu_1\ldots\mu_J}_{\nu_1\ldots\nu_J}(-1)^J \;,
\nn \\
&&-\int \frac{d\Omega}{4\pi} Sp\Big[\tilde V^{L>J}_{\mu_1\ldots\mu_J}(m_1+\hat k_1)
\tilde V^{L>J}_{\nu_1\ldots\nu_J}(m_2-\hat k_2)\Big]=
\nn \\
&&\qquad 2(s\!-\!\delta^2)|\vec k|^{2(J+1)}
O^{\mu_1\ldots\mu_J}_{\nu_1\ldots\nu_J}(-1)^J \ .
\label{loop_trip_1_fin}
\ee

\subsection{Cross sections}

The expressions for one-loop diagrams can be used to calculate
cross sections for different spin-orbital momentum states. The cross
section is given by
 \be
d\sigma&=&\frac{(2\pi)^4|A|^2}{4|\vec{q}|\sqrt{s}}\ d\Phi
        = \frac{|\vec{k}|}{|\vec{q}|}\ \frac{\rho(s)}{16\pi s}\
        \int\frac{d\Omega}{4\pi}|A|^2 \ .
 \ee
To calculate the amplitude squared $|A|^2$, one can use the expressions for
the one-loop diagram given by eqs. (\ref{loop_singlet}),
(\ref{loop_trip_1_fin}), (\ref{loop_trip_2}) and (\ref{loop_trip_3}).

For the $^1L_J$ state, one has
 \be
d\sigma=\frac{2J+1}{64\pi m^2}\ |\vec{k}|^{2J+1}\ |\vec{q}|^{2J-1}\ s\ .
 \ee
For the $^3L_J$ state $(L=J)$, the result is
 \be
d\sigma=\frac{2J+1}{64\pi m^2}\ |\vec{k}|^{2J+1}\ |\vec{q}|^{2J-1}\ s \ .
 \ee
The decay of the  $^3L_J$ states with
$(L<J)$ and $^3L_J\ (L>J)$  is determined by the sum of two
vertices:
\be
A_{tr}^{L\neq J} = \lambda_1 V^{L<J}_{\mu_1\ldots\mu_J} +
\lambda_2 V^{L>J}_{\mu_1\ldots\mu_J}
\ee
Then, the cross section is equal to
\be
d\sigma&=& \lambda_1^2 d\sigma_{11} + \lambda_2^2 d\sigma_{22}
+ \lambda_1\lambda_2 (d\sigma_{12} + d\sigma_{21})\ ,
\ee
where
 \be
d\sigma_{11}&=&\frac{2J+1}{256\pi s m^2}\ |\vec{k}|^{2J-1}\ |\vec{q}|^{2J-3}
\nn \\
&\times&
\left[2s-\frac{8J|\vec{k}|^2}{2J+1}\right]
\left[2s-\frac{8J|\vec{q}|^2}{2J+1}\right],
\nn \\
d\sigma_{22}
&=&\frac{2J+1}{256\pi s m^2}\ |\vec{k}|^{2J+3}\
|\vec{q}|^{2J+1} \nn \\ &\times&
\left[2s-\frac{8(J+1)|\vec{k}|^2}{2J+1}\right]
\left[2s-\frac{8(J+1)|\vec{q}|^2}{2J+1}\right],
\nn \\
d\sigma_{12}&=&\frac{1}{4\pi s m^2}\ \frac{J(J+1)}{2J+1}\
|\vec{k}|^{2J+3}\ |\vec{q}|^{2J+1},
\nn \\
d\sigma_{21}&=&\frac{1}{4\pi s m^2}\ \frac{J(J+1)}{2J+1}\
|\vec{q}|^{2J+3}\ |\vec{k}|^{2J+1}\ .
 \ee
For pure spin-1 operators, $V^{L<J}_{\mu_1\ldots\mu_J}$ and
$V^{L>J}_{\mu_1\ldots\mu_J}$, the cross section reads
\be
d\sigma&=& \lambda_1^2 d\sigma_{11} + \lambda_2^2 d\sigma_{22}\ ,
\ee
where
 \be
d\sigma_{11}&=&\frac{(2J+1)s}{64\pi m^2}\ |\vec{k}|^{2J-1}\ |\vec{q}|^{2J-3}\ ,
\nn \\
d\sigma_{22}&=&\frac{(2J+1)s}{64\pi m^2}\ |\vec{k}|^{2J+3}\ |\vec{q}|^{2J+1}\ .
 \ee


\section{ Decay into {\boldmath $3/2^+$} and {\boldmath $1/2^+$} particles}

Let $k_1$ be the momentum of the $3/2^+$ particle and $k_2$
 the momentum of $1/2^+$.
In this case, there are two spin states, $S=1$ and $S=2$.
Let us start from the $S=1$ states. Such  states
are constructed using the vector
spinors $\psi_\alpha$ for 3/2-spin particle and spin operators.

For $^3L_J$ $(J=L-1)$, that corresponds to \\$(0^-,1^+,2^-,3^+\ldots)$
states, and the operators read
 \be
W^{(1)}_{\mu_1\ldots\mu_J}& =&\bar \psi_\alpha(k_1)
V^{(1)\alpha}_{\mu_1\ldots\mu_J} u(-k_2)\ ,
\nn \\
V^{(1)\alpha}_{\mu_1\ldots\mu_J}& =& i \gamma_5
X^{(J+1)}_{\alpha\mu_1\ldots\mu_J} \ .
\label{operW1}
 \ee
For $^3L_J$ $(J=L+1)$, the operators are given by
 \be
W^{(2)}_{\mu_1\ldots\mu_J} & =&
\bar \psi_{\alpha_1}(k_1) i \gamma_5  X^{(J-1)}_{\alpha_2\ldots\alpha_J}
O^{\alpha_1 \ldots \alpha_J}_{\mu_1 \ldots \mu_J}   u(-k_2)\ ,
\nn \\
V^{(2)\alpha_1}_{\mu_1\ldots\mu_J} & =&
i \gamma_5  X^{(J-1)}_{\alpha_2\ldots\alpha_J}
O^{\alpha_1 \ldots \alpha_J}_{\mu_1 \ldots \mu_J} \ ,
\label{operW2}
 \ee
where the projection operator is needed for index symmetrisation.
For $^3L_J$ $(J=L)$ $(1^-,2^+,3^-,4^+\ldots)$\ , the operators
can be expressed as
 \be
W^{(3)}_{\mu_1\ldots\mu_J}\! & =&
 \gamma_5 \varepsilon_{\alpha_1\beta\xi\eta} \bar \psi_\beta(k_1)
k_\xi P_\eta X^{(J-1)}_{\alpha_2\ldots\alpha_J}
O^{\alpha_1 \ldots \alpha_J}_{\mu_1 \ldots \mu_J}
u(-k_2),                  \nn \\
V^{(3)\beta}_{\mu_1\ldots\mu_J} & =&
 \gamma_5 \varepsilon_{\alpha_1\beta\xi\eta}
k_\xi P_\eta X^{(J-1)}_{\alpha_2\ldots\alpha_J}
O^{\alpha_1 \ldots \alpha_J}_{\mu_1 \ldots \mu_J}\ .
\label{operW3}
 \ee

In case of $S=2$, there are five operators. For $^5L_J$ $(J=L+2)$ the operators
are given by
 \be
W^{(4)}_{\mu_1\ldots\mu_J} & =& \bar \psi_{\alpha_1}(k_1) \gamma_{\alpha_2}
O^{\alpha_1 \alpha_2}_{\nu_1 \nu_2}
X^{(J-2)}_{\nu_3 \ldots \nu_J}
O^{\nu_1 \ldots \nu_J}_{\mu_1 \ldots \mu_J}
u(-k_2),
\nn \\
V^{(4)\alpha_1}_{\mu_1\ldots\mu_J} & =&  \gamma_{\alpha_2}
O^{\alpha_1 \alpha_2}_{\nu_1 \nu_2}
X^{(J-2)}_{\nu_3 \ldots \nu_J}
O^{\nu_1 \ldots \nu_J}_{\mu_1 \ldots \mu_J}\ ,
 \ee
for $^5L_J$ $(J=L-2)$ by
 \be
W^{(5)}_{\mu_1\ldots\mu_J} & =& \bar\psi_{\alpha_1}(k_1) \gamma_{\alpha_2}
X^{(J+2)}_{\alpha_1\alpha_2\nu_1 \ldots \nu_J} u(-k_2),
\nn \\
V^{(5)\alpha_1}_{\mu_1\ldots\mu_J} & =& \gamma_{\alpha_2}
X^{(J+2)}_{\alpha_1\alpha_2\nu_1 \ldots \nu_J}\ ,
 \ee
for $^5L_J$ $(J=L)$ by
 \be
W^{(6)}_{\mu_1\ldots\mu_J} & =& \bar\psi_{\alpha}(k_1) \gamma_{\beta}
O^{\nu_1\xi}_{\alpha\beta}
X^{(J)}_{\xi\nu_2 \ldots \nu_J} O^{\nu_1 \ldots
\nu_J}_{\mu_1 \ldots \mu_J}
u(-k_2),
\nn \\
V^{(6)\alpha}_{\mu_1\ldots\mu_J} & =& \gamma_{\beta}
O^{\nu_1\xi}_{\alpha\beta}
X^{(J)}_{\xi\nu_2 \ldots \nu_J} O^{\nu_1 \ldots
\nu_J}_{\mu_1 \ldots \mu_J}\ ,
 \ee
for $^5L_J$ $(J=L-1)$ by
 \be
&&W^{(7)}_{\mu_1\ldots\mu_J}=
\nn \\
&&i \varepsilon_{\nu_1\beta\tau\eta}k_\tau P_\eta
O^{\alpha_1\alpha_2}_{\beta\xi}
\bar\psi_{\alpha_1}(k_1) \gamma_{\alpha_2}
X^{(J)}_{\xi\nu_2 \ldots \nu_J}
O^{\nu_1 \ldots \nu_J}_{\mu_1 \ldots \mu_J}
u(-k_2),
\nn \\
&&V^{(7)\alpha_1}_{\mu_1\ldots\mu_J}  =
i \varepsilon_{\nu_1\beta\tau\eta}k_\tau P_\eta
O^{\alpha_1\alpha_2}_{\beta\xi}
\gamma_{\alpha_2}
X^{(J)}_{\xi\nu_2 \ldots \nu_J}
O^{\nu_1 \ldots \nu_J}_{\mu_1 \ldots \mu_J}\ ,\ \qquad\;\;
 \ee
and for $^5L_J$ $(J=L+1)$ by
 \be
&&W^{(8)}_{\mu_1\ldots\mu_J} =
\nn \\
&&i \varepsilon_{\nu_1\beta\tau\eta}k_\tau P_\eta
O^{\alpha_1\alpha_2}_{\beta\nu_2}
\bar\psi_{\alpha_1}(k_1) \gamma_{\alpha_2}
X^{(J-2)}_{\nu_3 \ldots \nu_J}
O^{\nu_1 \ldots \nu_J}_{\mu_1 \ldots \mu_J}
u(-k_2),
\nn \\
&&V^{(8)\alpha_1}_{\mu_1\ldots\mu_J} =
i \varepsilon_{\nu_1\beta\tau\eta}k_\tau P_\eta
O^{\alpha_1\alpha_2}_{\beta\nu_2}
\gamma_{\alpha_2}
X^{(J-2)}_{\nu_3 \ldots \nu_J}
O^{\nu_1 \ldots \nu_J}_{\mu_1 \ldots \mu_J}.
\label{operW8}
 \ee

The one-loop diagram amplitudes for the corresponding operators are
calculated in Appendix 4.


\section{ Example: amplitude for the reaction \boldmath$pp \rightarrow
p K^+ \Lambda$\unboldmath}

Let us start from $pp$ scattering with the production of a
resonance $R$ in the intermediate state which decays into $K^+
\Lambda$. The diagram for the process is shown in Fig.
\ref{fig_example}.
\begin{figure}[h]
\leftline{
\epsfig{file=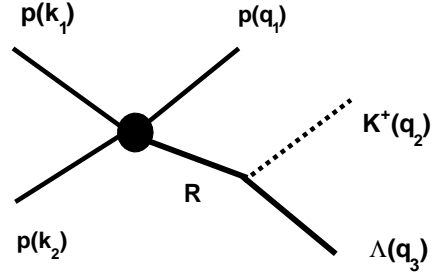,width=0.65\textwidth,clip=on}}
\caption{Reaction $pp \rightarrow p K^+ \Lambda$: $pp$ scattering with
production of a resonance $R$ in the intermediate state}
 \label{fig_example}
\end{figure}
Consider the partial wave amplitude for the $pp$ having quantum numbers
$J=n$, $L$ and $S$ in the initial state. The general form of the
angular dependent part of this partial amplitude is
\be
&& \Big( \bar u(-k_1)
Q^{(S,L,J)}_{\nu_1\ldots\nu_n} u(k_2) \Big)\; \bar u(q_3)
\tilde N_{\alpha_1\ldots\alpha_m}(R\!\to\!K \Lambda)
\nn \\
&&\times
F^{\alpha_1\ldots\alpha_m}_{\beta_1\ldots\beta_m}(q_2+q_3)
Q^{(S,L,J)}_{\beta_1\ldots\beta_m\nu_1\ldots\nu_n} u(-q_1)
\nn \\
&&-\{ k_1 \Leftrightarrow k_2 \}\ ,
\label{exam1}
\ee
where $P=q_1+q_2+q_3=k_1+k_2$.
The resonance $R$ with spin $J=m+1/2$ is produced in the intermediate
state and decays into a final state meson and a nucleon.

The initial $pp$ state operator $Q^{(S,L,J)}_{\nu_1\ldots\nu_n}$ is defined by
Eq. (\ref{V_1Lj}) for $S=0$ and  eqs. (\ref{V_3Lj_1_fin}),
(\ref{V_3Lj_2}), (\ref{V_3Lj_3}) for $S=1$. If the resonance in the
intermediate state has the spin $1/2\; (m=0)$, the same expressions
define the $R p$ state operator. For the spin-3/2 resonance in the
intermediate state, the operator
$Q^{(S,L,J)}_{\beta_1\ldots\beta_m\nu_1\ldots\nu_n}$ is defined by eqs.
(\ref{operW1})--(\ref{operW8}).
The operators for the $R\to 0^-\!+1/2^+$ transitions were
defined in \cite{decompos}:
\be
\hspace{-6mm}\tilde N^+_{\mu_1\ldots\mu_n}&=&X^{(n)}_{\mu_1\ldots\mu_n}
\qquad\quad\;\quad 1/2^-,3/2^+,5/2^-\ldots
 \nonumber\\
\hspace{-6mm}\tilde N^-_{\mu_1\ldots\mu_n}&=&i\gamma_\nu \gamma_5
X^{(n+1)}_{\nu\mu_1\ldots\mu_n}   \;\;\quad 1/2^+,3/2^-,5/2^+\ldots
\label{sum_v1}
\ee

Let us write down the amplitude for the $1/2^+$ resonance in the intermediate state.
In this case, one finds
\be
&&M = \sum_{S,L,J}\bar u(-k_1) Q^{(S,L,J)}_{\nu_1\ldots\nu_n} u(k_2)\;
\bar u(q_3)
\tilde N^-(q^\perp_{23})
\nonumber \\
&&
\times \frac{\hat q_2 +\hat q_3 + \sqrt{s_{23}}}{2\sqrt{s_{23}}}
 BW(s_{23})
Q^{(S,L,J)}_{\nu_1\ldots\nu_n} u(-q_1) A^{(S,L,J)}(s,s_{23})
\nonumber\\
&&\;\; -\;\;\{ k_1 \Leftrightarrow k_2 \} \ ,
\\
&&q^\perp_{23}=(q_2-q_3)_\nu
\Big ( g_{\mu\nu}-\frac{Q_{23\mu} Q_{23\nu}}{s_{23}}\Big ) ,
\quad Q_{23}=q_2+q_3\ ,\nn
\ee
where $A^{(S,L,J)}(s,s_{23})$ is the partial amplitude for
$pp\rightarrow R p$ and $BW(s_{23}$ parameterises the
resonance $R$).

Another type of processes, which may contribute to this reaction, is the
t-channel exchange of pseudoscalar and vector particles ($\pi$ and $\rho$).
This diagram is shown in Fig. \ref{fig_example1}.

\begin{figure}[h]
\leftline{
\epsfig{file=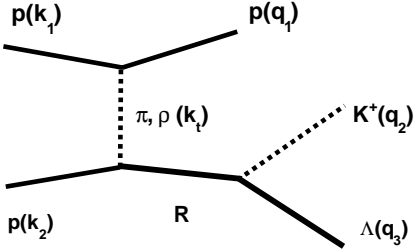,width=0.65\textwidth,clip=on}}
\caption{Reaction $pp \rightarrow p K^+ \Lambda$:
t-channel exchange diagram.}
\label{fig_example1}
\end{figure}

First, consider the exchange of a $\rho$ meson. The vertex operators for
the transition $baryon \rightarrow vector\;meson + baryon$ are given in
\cite{decompos}. Thus the $p \rho \rightarrow R$ operators are given by
\be
A_{lower}^{(i-)}&=&\bar u(Q_{23}) V^{(i-)\mu}(k_2^{\perp}) u(k_2)
\rho_\mu\ ,\qquad \qquad i=1,2 \ ,
\nn \\
k_{2\mu}^{\perp}&=&
\frac 12(k_2-k_t)_{\nu}
\Big ( g_{\mu\nu}-\frac{Q_{23\mu} Q_{23\nu}}{s_{23}}\Big ) \ ,
\ee
where
\be
V^{(1-)\mu}_{\alpha_1\ldots\alpha_{n}}(k^\perp)&=&
\gamma_\xi\gamma^\perp_\mu
X^{(n+1)}_{\xi\alpha_1\ldots\alpha_{n}}(k^\perp) \;,
\nn \\
V^{(2-)\mu}_{\alpha_1\ldots\alpha_{n}}(k^\perp)&=&
X^{(n+1)}_{\mu\alpha_1\ldots\alpha_{n}}(k^\perp) \;.
\label{vf_minus}
\ee
Here, $k_t = Q_{23}-k_2$  is the $\rho$-meson momentum.
In case of a spin-1/2 resonance in the intermediate state,
one should use Eq.(\ref{vf_minus}) with $n=0$. For the upper operator,
one has:
\be
A_{upper}^{(i-)}&=&\bar u(q_{1}) V^{(i-)\mu}(q_1^{\perp})
u(k_1) \rho_\mu\ ,\qquad \qquad i=1,2
\nn \\
q_{1\mu}^{\perp}&=& \frac
12(q_1-k_t)_{\nu} \Big ( g_{\mu\nu}-\frac{k_{1\mu}
k_{1\nu}}{k_{1}^2}\Big )
\ee
Summing over the polarisations yields:
\be
\sum\limits_{polarisations}\rho_\alpha\rho_\beta=
-g_{\alpha\beta}+\frac{k_{t\alpha}k_{t\beta}}{k_t^2}.
\ee

Finally, we arrive at the following amplitude for the $\rho$ exchange:
\be
 A^{(i)}_{\rho} &=& \bar u(q_{1}) V^{(i-)\mu}(q_1^{\perp}) u(k_1)
\bar u(q_3)\tilde N^-(q^\perp_{23})
\nn \\
&\times&
\frac{\hat q_2\! +\! \hat q_3\! +\! \sqrt
{s_{23}}}{2\sqrt{s_{23}}}
 BW(s_{23})
V^{(i-)\nu}(k_2^{\perp}) u(k_2)
\nn \\
&\times&
\Big(-g_{\mu\nu}\!+\!\frac{k_{t\mu}k_{t\nu}}{k_t^2}\Big) ,
\qquad i=1,2\ .
\label{x4a}
\ee
In case of  t-channel exchange of a pseudoscalar meson, $\pi$, one
should substitute the operator $V^{(i-)\mu}$ by $\tilde N^-$, so we
have \be
 A_{\pi} &=& \bar u(q_{1}) \tilde N^{-}(q_1^{\perp}) u(k_1)
\bar u(q_3)\tilde N^-(q^\perp_{23})\;
\nn \\ & \times&
\frac{\hat q_2\! +\! \hat q_3\! +\! \sqrt {s_{23}}}{2\sqrt{s_{23}}}
 BW(s_{23})
\tilde N^{-}(k_2^{\perp}) u(k_2)\ .
\label{x4a_1}
\ee

\section{Triangle-diagram amplitude with pion--nucleon
rescattering: logarithmic singularity}

In the amplitudes describing production of three-particle final states, the
unitarity condition is fulfilled automatically when final-state rescattering is
properly taken into account. However, rescattering may lead to singularities
where the amplitude tends to infinity. The triangle diagram with the $\Delta$
in the intermediate state gives us an example of this type of the process: it
has logarithmic singularity which under certain conditions ($\sqrt {s}\sim
m_N+m_\Delta$) can be near the physical region.

Because of $\sqrt {s}\sim m_N+m_\Delta$, we consider the amplitude
$pp\to N\Delta$ with $L'=0$ (the produced $N\Delta$ system is in the
$S$-wave). The quantum numbers of the final state are then restricted
to
 \be
J^P=1^+,2^+.
\ee
 The initial $pp$ system ($I=1$) has
\be
\hspace{-8mm}S=0\;: \quad L&=&0,2,4,...\quad J^P=0^+,2^+,4^+, ...\nn \\
\hspace{-8mm}S=1\;: \quad L&=&1,3,5,...\quad J^P=0^-,1^-,2^-,3^-, ...
\ee
We thus consider the transition
\begin{center}
$pp$ ($S$=0, $L$=2, $J^P$=$2^+$) $\to
N\Delta$ ($S'$=2, $L'$=0, $J^P$=$2^+$).
\end{center}
 The corresponding pole
amplitude reads:
\be \label{D18}
&&A^{pole}_{NN\to N N\pi}=
C^{pole}_{NN\to N N\pi}
 G^{(S=0,S'=2,L=2,L'=0,J=2)}_{pp\to
N\Delta}(s)
\nn \\
&&\times \left (\bar u(p'_1)g_\Delta k'^{\perp p_\Delta}_{1\mu}
\frac{\Delta_{\mu\nu}(p_{\Delta})}{m^2_\Delta-p^2_{\Delta}-im_\Delta
\Gamma_\Delta} \gamma_{\nu'} u(-p'_2) \right )
\nn \\
&&\times\left (\bar u(-p_2)
 i\gamma_5 X^{(2)}_{\nu\nu'}(k)
u (p_1)\right ).
\ee
Here, the factor  $C^{pole}_{NN\to N N\pi}$ refers to the isotopic
Clebsch-Gordan coefficients, and
\be \label{D18a}
k'^{\perp p_\Delta}_{1\mu}=
g^{\perp p_\Delta}_{\mu\mu'}p'_{1\mu'}.
\ee
The numerator of the $3/2-$spin fermion propagator is written in the
form used in \cite{ASB,decompos}:
\bea
\Delta_{\mu\nu}(k)&=&\frac{\hat k +M_\Delta }{2M_\Delta}
\Big(-g_{\mu\nu}^\perp +\frac 13
\gamma^\perp_\mu\gamma^\perp_\nu\Big), \nn \\
\gamma^\perp_\mu&=&g_{\mu\nu}^\perp \gamma_\nu \ , \quad
g_{\mu\nu}^\perp =g_{\mu\nu}-\frac{k_\mu k_\nu}{M_\Delta^2}.
\label{2-6}
\eea
The decay vertex $g_\Delta$ is determined by the imaginary part of the
loop diagram $\Delta \to N\pi \to \Delta$.
 For the sake of
simplicity, we change in (\ref{D18}):
$\Gamma_{\nu'}(k'_\perp)\to \gamma_{\nu'}$;
 however, using definition (\ref{purespinstate}) one can easily rewrite Eq.
(\ref{D18}) in a more rigid form.

Taking into account the rescattering process in the amplitude
(\ref{D18}), $\pi N\to \Delta \to \pi N$, one has
the following triangle-diagram amplitude (see Fig. \ref{fig_triangle}):
\be \label{D19}
&&A^{triangle}_{NN\to N N\pi}= C^{triangle}_{NN\to N N\pi}
 G^{(S=0,S'=2,L=2,L'=0,J=2)}_{pp\to N\Delta}(s)
\nn \\
&&\times
\Big (\bar u(p'_1)\Big[
\int \frac{d^4k_\pi}{i(2\pi)^4}\frac{1}{m^2_\pi-k^2_\pi-i0}
\nn \\
&&\times
g_\Delta k'^{\perp p'_\Delta}_{1\mu'}
\frac{\Delta_{\mu'\nu}(p'_{\Delta})}{m^2_\Delta-p'^2_{\Delta}-im_\Delta
\Gamma_\Delta} \gamma_{\nu'} \frac{-\hat p''_2+m}{m^2-p''^2_2-i0}
\nn \\
&&\times
g_\Delta k''^{\perp p''_\Delta}_{2\mu''}
\frac{\Delta_{\mu''\nu''}(-p_{\Delta})}{m^2_\Delta-p^2_{\Delta}-
im_\Delta \Gamma_\Delta}  g_\Delta k'^{\perp p_\Delta}_{2\nu''}
\Big ] u(-p'_2) \Big )
\nn \\
&&\times
\left (\bar u(-p_2) i\gamma_5 X^{(2)}_{\nu\nu'}(k) u (p_1)\right ).
\ee
Here,
\be \label{D19a}
k'^{\perp p'_\Delta}_{1\mu'}&=& g^{\perp p'_\Delta}_{\mu'\alpha}p'_{1\alpha} \ ,
\quad
k''^{\perp p''_\Delta}_{2\mu''}= g^{\perp p_\Delta}_{\mu''\alpha}p''_{2\alpha} \ ,
\nn \\
k'^{\perp p_\Delta}_{2\nu''}&=& g^{\perp p_\Delta}_{\nu''\alpha}p'_{2\alpha}\ ,
\ee
and
\be \label{D19b}
p'_\Delta &=&p'_1+k_\pi ,\quad p_{\Delta}=p'_2+p_\pi=p''_2+k_\pi\ ,
\nn \\
P&=&p'_\Delta +p''_2 \ .
\ee
\begin{figure}[h]
\vspace{-10mm}
\leftline{
\epsfig{file=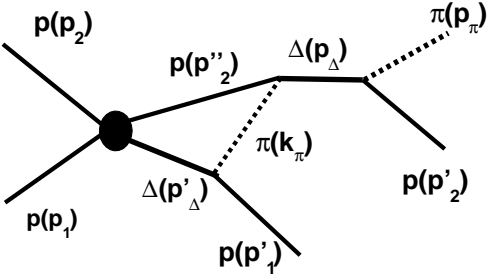,width=0.60\textwidth,clip=on}}
\caption{Triangle diagram with final state pion--nucleon rescattering}
\label{fig_triangle}
\end{figure}

One can simplify (\ref{D19}) by extracting the numerator in the
singular point that corresponds to
\be \label{D19c}
m^2_\Delta=p'^2_{\Delta}\ , \qquad
m^2=p''^2_2\ ,\qquad m^2_\pi=k^2_\pi\ .
\ee
Then, Eq. (\ref{D19}) reads:
\be
&&A^{triangle}_{NN\to N N\pi}= C^{triangle}_{NN\to N N\pi}
 G^{(S=0,S'=2,L=2,L'=0,J=2)}_{pp\to N\Delta}(s)
\nn \\ &&
\times\Big (\bar u(p'_1) g_\Delta k'^{\perp p'_\Delta}_{1\mu'}({\rm tr})
\Delta_{\mu'\nu}(p'_{\Delta}({\rm tr})) \gamma_{\nu'}
(-\hat p''_2({\rm tr})+m) g_\Delta
\nn \\ &&
\times k''^{\perp p''_\Delta}_{2\mu''}({\rm tr})
\frac{\Delta_{\mu''\nu''}(-p_{\Delta})}{m^2_\Delta-p^2_{\Delta}-
i m_\Delta \Gamma_\Delta}  g_\Delta k'^{\perp p_\Delta}_{2\nu''}
 u(-p'_2) \Big )
\nn \\ &&
\times\left (\bar u(-p_2)
 i\gamma_5 X^{(2)}_{\nu\nu'}(k)
u (p_1)\right )
\nn \\ &&\times\int \frac{d^4k_\pi}{i(2\pi)^4}\frac{1}{m^2_\pi-k^2_\pi-i0}
\frac{1}{m^2-(p_\Delta-k_\pi)^2-i0}
 \nn \\ &&
\times \frac{1}{m^2_\Delta-(P-p_{\Delta}+k_\pi)^2-im_\Delta \Gamma_\Delta}
\label{D20}
\ee
The momenta
$k'^{\perp p'_\Delta}_{1\mu'}({\rm tr})$,
$p'_{\Delta}({\rm tr})$,
$p''_2({\rm tr})$,
$k''^{\perp p''_\Delta}_{2\mu''}({\rm tr})$
obey the constraints (\ref{D19c}). The integral in (\ref{D20})
corresponds to the triangle diagram with spinless particles. Its
calculation is performed in Appendix 5.


\section{Box-diagram singularities in
the reaction \boldmath$NN\to \Delta\Delta\to NN\pi\pi$\unboldmath}

The primary aim of a partial wave analysis is to extract the pole
singularities of amplitudes, thus determining resonances. Of course,
the existence of other singulari\-ti\-es like thre\-shold singularities
should be taken into account. This is possible using the $K$-matrix
technique, see \cite{Kmatrix,K1,K2} and references therein.
Singularities due to resonances in the intermediate state need more
sophisticated treatment.

The existence of triangle-diagram singulari\-ti\-es, which may be
located near the physical region of a three-particle production
reaction, was proven in \cite{aitchison,ad}: these singularities
diverge as $\ln (s-s_0)$. Stronger singularities (with a
$(s-s_0)^{-1/2}$ behaviour) are related to box diagrams
\cite{vva,norton}.

Here, we present box-diagram and triangle-diagram singular
amplitudes for the reaction $NN\!\to\!\Delta\Delta\!\to\! NN\pi\pi$
taking into account the spin structure in a way which allows us to
include these singular amplitudes into partial wave analyses (this was
not yet done in \cite{vva,norton}).

Let us introduce the following notations for the two-pole and
box diagrams in the reactions $NN\to\Delta\Delta\to NN\pi\pi$
(see Figs. \ref{fig_poleDD} and \ref{fig_box1}).

The initial state momenta are:
\be  \label{2-1}
P_1+P_2=P, \quad P^2=W^2,\quad \frac 12 (P_1-P_2)=q
\ee
Final state momenta:
\bea \label{2-2}
&&(p_1+p_3)^2=s_{13}, \quad (p_1+p_3+p_2)^2=s_{4},\nn \\
&& p_1+p_3 =k_1,\quad  k_1^\perp =\frac 12 (p_1-p_3)^{\perp k_1}=
      p_1^{\perp k_1}=- p_3^{\perp k_1},\nn \\
&&(p_2+p_4)^2=s_{24}, \quad (p_2+p_4+p_1)^2=s_{1}, \nn \\
&& p_2+p_4 =k_2,\quad  k_2^\perp=\frac 12 (p_2-p_4)^{\perp k_2}=
p_2^{\perp k_2}=-p_4^{\perp k_2} , \nn \\
&&
(p_1+p_2)^2=s, \quad p_1+p_2=p\, .
\eea
Here, the symbol $\perp$$k_i$ means the component of
a vector  perpendicular to $ k_i$:
\be \label{2-3}
p_\mu^{\perp k_i} =p_\mu - k_{i\mu} \frac {(k_ip)}{k_i^2}\, .
\ee

\subsection{\boldmath$(NN)_{S-wave}$\unboldmath\  state with
\boldmath$J^{P}=0^+$\unboldmath , two-pole diagram}

In $pp$ collision with $I=1$, the $S$-wave $\Delta\Delta$
state is produced. First, consider the two-pole diagram of
 Fig. \ref{fig_poleDD}.
\begin{figure}[h]
\leftline{
\epsfig{file=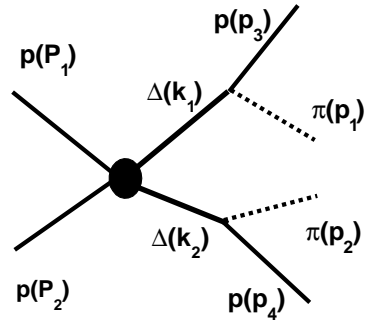,width=0.60\textwidth,clip=on}}
\caption{Pole diagram for reaction
$NN\to \Delta\Delta\to NN\pi\pi$}
\label{fig_poleDD}
\end{figure}
The amplitude for the production and decay of two $\Delta$-isobars, $NN\to
\Delta\Delta\to NN\pi\pi$, omitting charge indices and
corresponding  Clebsch-Gordan coefficients, reads:
\be
&&A_{NN\to \Delta\Delta\to (N\pi)(N\pi)}= \Big( \bar u(-P_2)
u (P_1)\Big ) G_{NN\to \Delta\Delta}(W) \nn \\
&&\times \Big (\bar u(p_3)g_\Delta k_{1\mu}^\perp
\frac{\Delta_{\mu\nu'}(k_1)}{M^2_\Delta-s_{13}-iM_\Delta \Gamma_\Delta}
\nn \\ &&
 \times\frac{\Delta_{\nu'\nu}(-k_2)}{M^2_\Delta-s_{24}-iM_\Delta
\Gamma_\Delta}
(-) k_{2\nu}^\perp g_\Delta u(-p_4)
\Big).
\label{2-4}
\ee

\subsection{Box-diagram amplitude with pion-pion rescattering}

The box-diagram amplitude with pion-pion rescattering in the Feynman
technique  (see Fig. \ref{fig_box1}) is equal to
\begin{figure}[h]
\leftline{
\epsfig{file=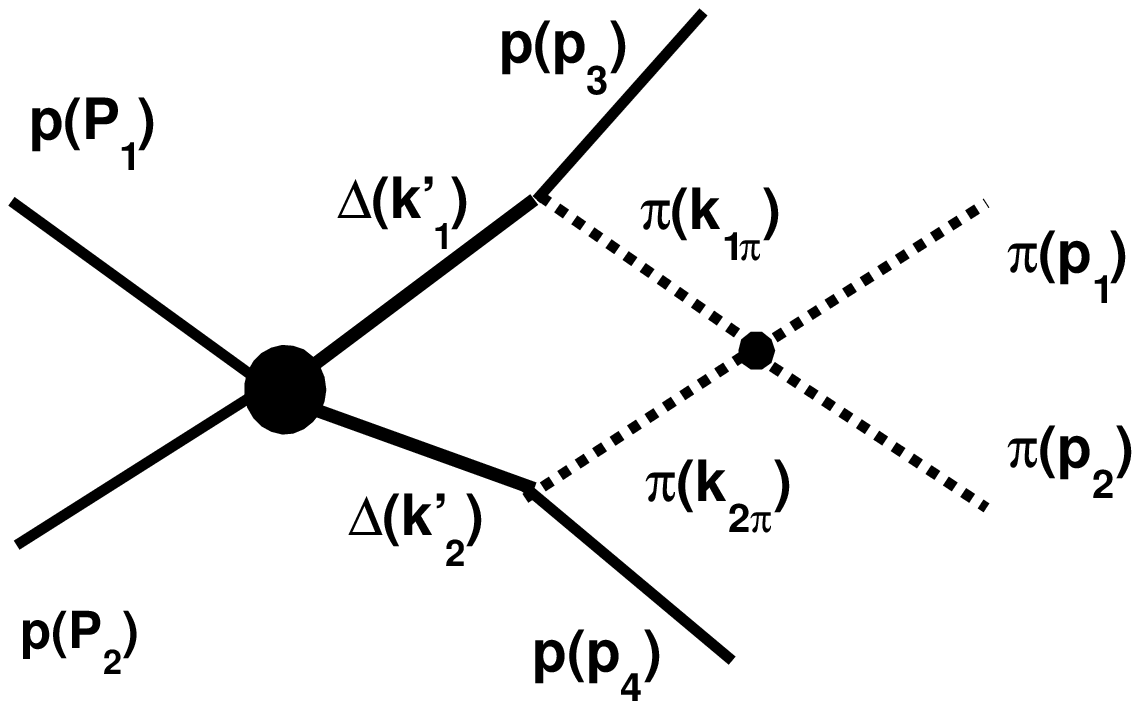,width=0.60\textwidth,clip=on}}
\caption{Box-diagram with pion-pion rescattering}
\label{fig_box1}
\end{figure}
\be
\label{2-7}
&&A_{NN\to \Delta\Delta\to NN +(\pi\pi\to \pi\pi)_S}=
A^{S-wave}_{\pi\pi\to\pi\pi}(s)
\nn \\ &\times&
G_{NN\to \Delta\Delta}(W)
\Big(\bar u(-P_2) u(P_1)\Big )
 \Big( \bar u(p_3)\Big [
\int\frac{d^4k'}{i(2\pi)^4}
\nn \\ &\times&
\frac{ g_{\Delta} k'^{\perp}_{1\mu} \Delta_{\mu\nu'}(k'_1)
\Delta_{\nu'\nu}(-k'_2)(-) k'^\perp_{2\nu} g_\Delta}
{(M^2_{\Delta}-s'_{13}-iM_{\Delta} \Gamma_{\Delta})
(M^2_\Delta-s'_{24}-iM_\Delta \Gamma_\Delta   )}
\nn \\ &\times&
\frac{1}{(m^2_{\pi}-k^2_{1\pi}-i0)(m^2_{\pi}-k^2_{2\pi}-i0)}\Big ]
u(-p_4) \Big).
\ee
The factor $ A^{S-wave}_{\pi\pi\to\pi\pi}(s)$ is the
$S$-wave $\pi\pi$-scattering
amplitude. Here we take into account the low-energy
$\pi\pi$ interaction only. In the $K$-matrix representation, it is
written in the form \be
 A^{S-wave}_{\pi\pi\to\pi\pi}(s)=
\frac{K(s)}{1-i\rho(s)K(s)}, \;\;
\rho(s)=\frac{1}{16\pi}
\sqrt{\frac{s-4m^2_{\pi}}{s} }  \, . \nn \\
\label{2-7a}
\ee
In (\ref{2-7a}), we take into account the full $S$-wave as observed
experimentally, including the so-called sigma-meson, independently of
its existence. Generally speaking, it is possible to account for higher
waves as well, but the box diagram with two $\Delta$'s leads to
singularities near the physical region of the production process at
$\sqrt s \la 0.6$ GeV only.

The approximation used in the calculation of the box diagram
(\ref{2-7}) is related to  the extraction of the leading terms of the
singular amplitude. To this aim, we fix the numerator of the integrand
in the propagator poles by setting
\be  \label{2-8}
\hspace{-2mm}k'^2_1\to
M^2_\Delta\, ,\;
  k'^2_2\to M^2_\Delta\, ,\;
k^2_{1\pi}\to m^2_{\pi},  \;
k^2_{2\pi}\to m^2_{\pi},\qquad
\ee
which leads in (\ref{2-7}) to the substitution
\bea
k'^{\perp}_{1\mu} \to  k^{\perp}_{1\mu}({\rm box})&=&-
 p_3^{\perp k_1({\rm box})}, \;
 k'_1 \to k_1({\rm box}), \nn \\
 k'^\perp_{2\nu}\to k^\perp_{2\nu}({\rm box})&=&-
p_4^{\perp k_2({\rm box})}, \;
k'_2\to k_2({\rm box})). \qquad
\eea
Now, in the c.m. system, the momenta $ k_a({\rm box})$ read
\be
 k_1({\rm box})&=&( W/2 , 0,0, \sqrt {W^2/4-M^2_\Delta}), \nn \\
 k_2({\rm box})&=&( W/2 , 0,0,- \sqrt {W^2/4-M^2_\Delta}).
\ee
Here, we denote the four-momentum as $k=(k_0,\!k_x,\!k_y,\!k_z)$.
Under the constraints of Eq. (\ref{2-8}), the numerator of the
integrand does not depend on the integration variables, and it can be
written separately for the leading singular ($LS$) term:
\be
&&
A^{(LS)}_{NN\to \Delta\Delta\to NN +(\pi\pi\to \pi\pi)_S}=
A^{S-wave}_{\pi\pi\to\pi\pi}(s)
G_{NN\to \Delta\Delta}(W)
\nn \\
&&\times
\Big (\bar u(-P_2) u(P_1)\Big )
 \Big (\bar u(p_3)
 g_{\Delta}(- p^{\perp k_1({\rm box})}_{3\mu})
 \Delta_{\mu\nu'}(k_1({\rm box}))
\nn \\
&&\times
\Delta_{\nu'\nu}(-k_2({\rm box})) p^{\perp k_({\rm box})}_{4\nu}
g_\Delta
 u(-p_4)\Big )
\nn \\
&&\times
\int\frac{d^4k'}{i(2\pi)^4}
\frac{ 1}
{(M^2_{\Delta}-(\frac 12 p+k'+p_3)^2-iM_{\Delta} \Gamma_{\Delta})}
\nn \\
&&\times
\frac{1}{
(M^2_\Delta-(\frac 12 p-k'+p_4)^2-iM_\Delta \Gamma_\Delta   )}
\nn \\
&&\times
\frac{1}{
(m^2_{\pi}-(\frac 12 p+k')^2-i0)(m^2_{\pi}-(\frac 12 p-k')^2-i0)} \ , \qquad
\label{2-9}
\ee
where
\be  \label{2-9a}
\frac 12 p+k' =k_{1\pi}, \quad \frac 12 p-k' =k_{2\pi}
\ ,\quad p_1+p_2=p\ .\qquad
\ee
The box-diagram integral of Eq. (\ref{2-9}) is calculated in Appendix
6: in this Appendix, we demonstrate the effects of the box diagram on
the $\pi\pi$ spectra.

\subsection{Box-diagram amplitude with pion--nucleon rescattering}

In the Feynman technique, the box-diagram amplitude with
pion--nucleon rescattering in the resonance state $(I=3/2, J=3/2)$
reads (see Fig. \ref{fig_box2}):
\begin{figure}[h]
\leftline{
\epsfig{file=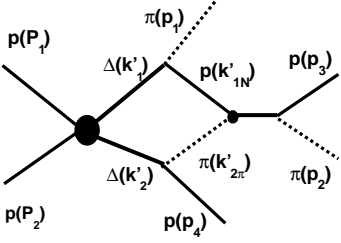,width=0.45\textwidth,clip=on}}
\caption{Box-diagram with pion--nucleon rescattering}
\label{fig_box2}
\end{figure}
\be
&&
A_{NN\to \Delta\Delta\to N\pi +(N\pi\to N\pi)_{\Delta} }=
G_{NN\to \Delta\Delta}(W)
\nn \\ &&
\times \Big (\bar u(-P_2) u(P_1)\Big )
\nn \\ &&
\times
\Big(\bar u(p_3) g_{\Delta}\frac 12 (p_{2}-p_{3})^{\perp p_{\Delta}}_\mu
\frac{\Delta_{\mu\mu'}(p_{\Delta})}{M^2_{\Delta}- p^2_{\Delta}-iM_{\Delta}\Gamma_\Delta}
\nn \\
&&\times
\Big [\int\frac{d^4k'}{i(2\pi)^4} \frac 12 (k'_{2\pi}-k'_{1N})^{\perp p_{\Delta}}_{\mu'}
g_{\Delta} \frac{\hat k'_{1N} +m_N}{m_N^2-k'^2_{1N}-i0}  g_{\Delta}
\nn \\
&& \times
\frac{\frac 12
(p_1-k'_{1N})^{\perp k'_1}_{\mu'}
\Delta_{\mu'\nu'}(k'_1)
\Delta_{\nu'\nu}(-k'_2)\frac 12 (- k'_{2\pi}+p_4)^{\perp k'_2}_{\nu}}
{(M^2_{\Delta}-k'^2_1-iM_{\Delta} \Gamma_{\Delta})
(M^2_\Delta-k'^2_2-iM_\Delta \Gamma_\Delta   )}
\nn \\
&&\times\frac{1}{(m^2_{\pi}-k'^2_{2\pi}-i0)}
\Big ]\,g_\Delta u(-p_4)\Big]\ ,
\label{2-10}
\ee
where $p_{\Delta}=p_2+p_3$.
By fixing the numerator of (\ref{2-10}) at
\be  \label{2-10b}
k'^2_1\to M^2_\Delta\, ,\; k'^2_2 \to M^2_\Delta\, ,\;
k^2_{1\pi}\to m^2_{\pi}, \;   k^2_{1N} \to m^2_{N},\qquad
\ee
we write
the leading singular ($LS$) terms of the box-dia\-gram amplitude
as follows:
\be
\label{2-11}
&&
A^{(LS)}_{NN\to
\Delta\Delta\to N\pi +(N\pi\to N\pi)_{\Delta} }=
G_{NN\to \Delta\Delta}(W)
\nn \\ &&\times
\Big (\bar u(-P_2) u (P_1)\Big )
\nn \\ &&\times
\Big (\bar u(p_3)
 g_{\Delta}
\frac 12 \large(p_{2}-p_{3}\large )^{\perp p_{\Delta}}_\mu
\frac{\Delta_{\mu\mu'}(p_{\Delta})}{M^2_{\Delta}- p^2_{\Delta}-iM_{\Delta}\Gamma_\Delta}
\nn \\ &&\times
\frac 12 \large(k_1({\rm box})-p_1-k_2({\rm box})+p_4
\large )^{\perp p_{\Delta}}_{\mu'}
\nn \\ && \times
g_{\Delta}
 \left (\hat k_1({\rm box})-\hat p_1 +m_N\right )\, g_{\Delta} \,
 p^{\perp k_1({\rm box})}_{1\mu'}
\nn \\ && \times
\Delta_{\mu'\nu'}(k_1({\rm box}))
\Delta_{\nu'\nu} (-k_2({\rm box})
)p^{\perp k_2({\rm box})}_{4\nu} \,
g_\Delta\, u(-p_4)
\Big )
\nn \\ &&
   \times
\int\frac{d^4k_\pi}{i(2\pi)^4}
 \frac{1}{(m_N^2-(p_{\Delta}-k^2_{\pi})^2-i0)}
\nn \\ &&\times
\frac{1}{(M^2_{\Delta}-(p_{\Delta}-k_{\pi}+p_1)^2-iM_{\Delta}\Gamma_{\Delta})}
\nn \\ &&
\times
\frac {1} { (M^2_\Delta-(k_\pi+p_4)^2-iM_\Delta \Gamma_\Delta
)(m^2_{\pi}-k^2_{\pi}-i0)} \ .
\ee

\subsection{\boldmath$(NN)_{D-wave}$\unboldmath\  state with
\boldmath$J^{P}=2^+$\unboldmath, two-pole and box diagrams}

The production of $\Delta\Delta$ near the threshold in the S-wave
leads to a $J^{P}=2^+$ state as well and, correspondingly, to a strong
box-diagram singularity in this wave. In the $J^{P}=2^+$ wave, the
transition $(NN)_{D-wave}\to (\Delta\Delta)_{S-wave}$ is related to the
two-pole amplitude
\be
&&A_{(NN)_D\to (\Delta\Delta)_S\to (N\pi)(N\pi)}=
G_{NN\to \Delta\Delta}(W)
\nn \\
&&\times
\left (\bar u(-P_2)X^{(2)}_{\nu'\nu''}(q) u (P_1)\right )
\nn \\
&&\times
\Big (\bar u(p_3)g_\Delta k_{1\mu}^\perp
\frac{\Delta_{\mu\nu'}(k_1)}{M^2_\Delta-s_{13}-iM_\Delta \Gamma_\Delta}
\nn \\ &&\times
\frac{\Delta_{\nu''\nu}(-k_2)}{M^2_\Delta-s_{24}-iM_\Delta \Gamma_\Delta}
(- k_{2\nu}^\perp) g_\Delta u(-p_4)
\Big )\ .
\label{3-4}
\eea
The box-diagram amplitude with the  pion--pion rescattering is given by
\be
&&
A^{(LS)}_{NN\to \Delta\Delta\to NN +(\pi\pi\to \pi\pi)_S}=
A^{S-wave}_{\pi\pi\to\pi\pi}(s)
 G_{NN\to \Delta\Delta}(W)
\nn \\ &&\times
\left (\bar u(-P_2)X^{(2)}_{\nu'\nu''}(q) u (P_1)\right )\,
\nn \\ &&\times
 \Big (\bar u(p_3)
 g_{\Delta}(- p^{\perp k_1({\rm box})}_{3\mu})
 \Delta_{\mu\nu'}(k_1({\rm box}))
\nn \\ &&\qquad\times
\Delta_{\nu''\nu}(-k_2({\rm box})) p^{\perp k_({\rm box})}_{4\nu}
g_\Delta
 u(-p_4)\Big )
\nn \\ &&\times
\int\frac{d^4k'}{i(2\pi)^4}
\frac{ 1}
{(M^2_{\Delta}-(\frac 12 p+k'+p_3)^2-iM_{\Delta} \Gamma_{\Delta})}
\nn \\ &&\times
\frac{1}{(M^2_\Delta-(\frac 12 p-k'+p_4)^2-iM_\Delta \Gamma_\Delta   )}
\nn \\ &&\times
\frac{1}{
(m^2_{\pi}-(\frac 12 p+k')^2-i0)(m^2_{\pi}-(\frac 12 p-k')^2-i0)}\ .\qquad
\label{3-9}
\ee
In the leading singular-term approach, the box-diagram amplitude with
the pion--nucleon rescattering can be written in the form
\be &&
A^{(LS)}_{NN\to \Delta\Delta\to N\pi +(N\pi\to N\pi)_{\Delta} }=
 G_{NN\to \Delta\Delta}(W)
\nn \\ &&\times
\left (\bar u(-P_2) X^{(2)}_{\nu'\nu''}(q) u (P_1)\right )
\nn \\ &&\times
\Big (\bar u(p_3)
 g_{\Delta}\frac 12 \large(p_{2}-p_{3}\large )^{\perp p_{\Delta}}_\mu
\frac{\Delta_{\mu\mu'}(p_{\Delta})}{M^2_{\Delta}-
 p^2_{\Delta}-iM_{\Delta}\Gamma_\Delta}
\nn \\ &&\times
\frac 12 \large(k_1({\rm box})-p_1-k_2({\rm box})+p_4
\large )^{\perp p_{\Delta}}_{\mu'}
\, g_{\Delta}
\nn \\ &&\times
\left (\hat k_1({\rm box})-\hat p_1 +m_N\right )
 g_{\Delta} \,
 p^{\perp k_1({\rm box})}_{1\mu'}
\Delta_{\mu'\nu'}(k_1({\rm box}))
\nn \\ &&\times
\Delta_{\nu''\nu} (-k_2({\rm box})
)p^{\perp k_2({\rm box})}_{4\nu} \,
g_\Delta\, u(-p_4)
\Big )
\nn \\ &&\times
\int\frac{d^4k_\pi}{i(2\pi)^4}
 \frac{1}{(m_N^2-(p_{\Delta}-k^2_{\pi})^2-i0)}
\nn \\ &&\times
\frac{1}{(M^2_{\Delta}-(p_{\Delta}-k_{\pi}+p_1)^2-iM_{\Delta}\Gamma_{\Delta})}
\nn \\ &&\times
\frac {1} { (M^2_\Delta-(k_\pi+p_4)^2-iM_\Delta \Gamma_\Delta
)(m^2_{\pi}-k^2_{\pi}-i0)} \ .\qquad\quad
\label{3-11}
\ee


\section{Conclusion}

We have developed a new method for the partial wave analysis of
data on the baryon-baryon and baryon-anti\-baryon collision.
The method is based on the
 operator decomposition approach which was successfully
applied before to a number of me\-son-induced reactions. The
article emphasises the analysis of reactions with three or four
particles in the final state, where triangle and box singularities
might play an important role. A full set of partial wave
amplitudes is constructed for nucleon--nucleon elastic scattering and
for $N\Delta$ and $\Delta\Delta$ production. With these amplitudes,
expressions for partial widths and for reaction cross sections are
presented. Some examples how to calculate contributions from triangle
and box diagrams in simple cases are explicitly given. The application
of the methods developed here to the analysis of new data obtained
and expected from COSY should provide valuable information about the
hadron spectrum and properties of hadron interaction.

\section*{Acknowledgments}

We would like to thank  L.G. Dakhno for helpful discussi\-ons and
critical reading of the manuscript. The work was supported by a FFE
grant of the Research Center J\"ulich and by the Deutsche
Forschungsgemeinschaft within the Sonderforschungsbereich SFB/TR16.
We$\,$ would$\,$ like to thank the Alexander von Humboldt foundation for
generous support in the initial phase of the project, A.V.A. for a AvH
fellowship and A.V.S. for the Friedrich-Wilhelm Bessel award.
A.~Sarantsev gratefully acknowledges the support from Russian Science
Support Foundation. This work is also supported by Russian Foundation
for Basic Research
 and RSGSS 5788.2006.2 (Russian State Grant Scientific School).


\section* {Appendix 1}

The baryon wave functions $\psi (p)$ and
$\bar \psi(p)=\psi^+(p)\gamma_0$  obey the Dirac equation
\begin{equation} \label{TB1}
(\hat p-m)\psi(p)=0\ , \qquad  \bar \psi(p)(\hat p-m)=0\ .
\end{equation}
The $\gamma$-matrices were used in the form
\begin{eqnarray} \label{TB2}
&& \gamma_0 \ =\left(\begin{array}{cc} I&0 \\ 0&-I\end{array} \right),
\quad \bg \ =\left(\begin{array}{cc} 0&\bs \\ -\bs&0\end{array} \right),
\nn \\
&& \gamma_5\ =\ i\gamma_0\gamma_1\gamma_2\gamma_3=\
\left({0\ I \atop I\ 0} \right)\;,
\nn \\
&& \gamma_0^+=\gamma_0\ , \quad \bg^+=-\bg\ ,
\end{eqnarray}
with the standard Pauli matrices
\begin{eqnarray}\label{TB3}
&&\sigma_1=\left( \begin{array}{cc}
0 & 1 \\ 1 & 0 \end{array} \right), \;
\sigma_2=\left(\begin{array}{cc} 0 &-i \\ i & 0 \end{array}
\right), \;
\sigma_3=\left(\begin{array}{cc} 1&0\\ 0&-1\end{array} \right),\nn \\
&& \sigma_a\sigma_b\ =\ i\varepsilon_{abc}\sigma_c\ .
\end{eqnarray}
The Dirac equation gives four wave functions 
\begin{eqnarray}\label{TB4}
j=1,2: &&\;\;
\psi_j(p) = \sqrt{p_0+m}\left(\begin{array}{c}
\varphi_j \\ \frac{(\mbox{\boldmath$\sigma p$})}{p_0+m}\ \varphi_j
\end{array}\right), \\&&
\;\; \bar \psi_j(p)=
\sqrt{p_0+m}\left(\varphi^+_j, -\varphi^+_j \frac{
(\mbox{\boldmath$\sigma p$})}{p_0+m}\right) \nonumber
\\\label{TB4a}
j=3,4:&& \;\; \psi_j(-p) =i\sqrt{p_0+m} \left(
\begin{array}{c} \frac{(\mbox{\boldmath$\sigma p$})}{p_0+m}\ \chi_j \\
\chi_j \end{array}\right),\\&&
 \;\; \bar \psi_j(-p) = -i\sqrt{p_0+m}
\left(\chi^+_j  \frac{ (\mbox{\boldmath$\sigma p$})}{p_0+m}
\,,-\chi^+_j\right), \nonumber
\end{eqnarray}
where $\varphi_j$ and $\chi_j$ are two-component spinors
\be  \label{TB5}
\varphi_j=
\left(\begin{array}{c}\varphi_{j1} \\\varphi_{j2}
\end{array} \right), \qquad
\chi_j=
\left(\begin{array}{c}\chi_{j1}\\ \chi_{j2}
\end{array} \right),
\ee
which are normalized as follows:
\be \label{TB6}
\varphi^+_j \varphi_\ell =\delta_{j\ell}, \quad  \chi^+_j \chi_\ell
=\delta_{j\ell}\ .
\ee
The solutions with $j=3,4$ refer to antibaryons. The corresponding wave
function is given by
\be
j=3,4&:&\qquad
\psi^c_j(p)\ =\ C\bar \psi^T_j(-p),\label{TB7}
\ee
where
\be \label{TB8}
 C=\gamma_2\gamma_0=
\left(\begin{array}{cc}
0 & -\sigma_2\\
-\sigma_2 & 0\end{array}\right)
=\left(\begin{array}{cccc}
0 & 0 & 0 & i\\
0 & 0 & -i & 0\\
0 & i & 0 & 0 \\
-i & 0 & 0 & 0 \end{array} \right)\ .
\ee
The equation (\ref{TB7}) reads:
\be
\label{TB9}
&&j = 3,4:
\\
&&\psi^c_j(p)=\left(\begin{array}{c} 0\ -\sigma_2\\ -\sigma_2\ 0 \end{array}
\right) i \sqrt{p_0+m} \left( \begin{array}{c}
-\frac{(\mbox{\boldmath$\sigma^{T}p$})}{p_0+m}\ \chi^*_j\\
 \chi^*_j \end{array} \right)=
\nn \\
&&-i\sqrt{p_0+m}
\left( \begin{array}{c} \sigma_2\chi^*_j \\
\frac{(\mbox{\boldmath$\sigma p$})}{p_0+m}\ \sigma_2\chi^*_j
\end{array}\right)
=\sqrt{p_0+m} \left(\begin{array}{c} \varphi^c_j\\
\frac{(\mbox{\boldmath$\sigma p$})}{p_0+m}\ \varphi^c_j
\end{array}\right) . \nn
\ee
In (\ref{TB9}), we have used the commutator
\be
\sigma_2(\mbox{\boldmath$\sigma^{T}p$})=-\sigma_1p_1\sigma_2
-\sigma_2p_2\sigma_2-\sigma_3p_3\sigma_2=-(\mbox{\boldmath$\sigma p$})
\sigma_2. \nn \\
\ee
We define the two-component spinor for anti\-bary\-ons as
\be \label{TB10}
\varphi^c_j=-i\sigma_2\chi^*_j=\left(\begin{array}{cc}
0 & -1 \\ 1 & 0 \end{array}\right)\chi^*_j=
\left(\begin{array}{c} -\chi^*_{j2}\\ \chi^*_{j\ell}
\end{array} \right).
\ee
The wave functions defined in eqs. (\ref{TB4})-(\ref{TB4a}) are
normalized as follows:
\begin{eqnarray} \label{TB11}
 j,\ell=1,2&:&\quad \left(\bar \psi_j(p)\psi_\ell(p)\right)= 2m\ \delta_{j\ell},
\nn \\
j,\ell=3,4&:&\quad \left(\bar \psi_j(-p)\psi_\ell(-p)\right)=-2m\
\delta_{j\ell}. \end{eqnarray}
They obey the completeness relation
\begin{eqnarray} \label{TB12}
&& \sum_{j=1,2}\, \psi_{j\alpha}(p)\,\bar \psi_{j\beta}(p)\ =\
(\hat p+m)_{\alpha\beta}\ ,
\nn \\
&& \sum_{j=3,4}\, \psi_{j\alpha}(-p)\, \bar \psi_{j\beta}(-p)\ =\
(\hat p-m)_{\alpha\beta}\ .
\end{eqnarray}

\section* {Appendix 2}

{\bf (i) The $S$-wave terms in the  the nonrelativistic limit.}
\\ We consider the operators with $L=0$ from eqs. (\ref{V_1Lj}) and (\ref{V_3Lj_2pure})
in the c.m. system
(${\bf p}_1=-{\bf p}_2={\bf k}$ and ${\bf p'}_1=-{\bf p'}_2={\bf k'}$).
For $L=0$, we have
the following operators in the nonrelativistic approach:
\be
Q^{000}(k) =i\gamma_5= i
\left(\begin{array}{cc}0 & I\\ I & 0 \end{array} \right), \nn \\
Q^{101}(k) =\Gamma_\mu\simeq
\left(\begin{array}{cc}0 & \mbox{\boldmath$\sigma$}\\
-\mbox{\boldmath$\sigma$} & 0 \end{array} \right).
\label{TB27}
\ee
In the c.m. system, we have:
\begin{eqnarray}
\label{TB28}
j, j'=1,2&:&
\psi_{Nj}(p_1) \simeq \sqrt{2m_N}\left(\begin{array}{c}
\varphi_{Nj} \\
\frac{(\mbox{\boldmath$\sigma k$})}{2m_N}\ \varphi_{Nj}
\end{array}\right), \\ &&
 \bar \psi_{Nj'}(p'_1)\simeq
\sqrt{2m_N}\left(\varphi^+_{Nj'}, -\varphi^+_{Nj'}
\frac{(\mbox{\boldmath$\sigma k'$})}{2m_N}\right),
\nn
\\
\ell,\ell '=3,4&:&  \psi^c_{\Lambda \ell '}(-p'_2) \simeq
i\sqrt{2m_\Lambda}
 \left(
\begin{array}{c} \frac{-(\mbox{\boldmath$\sigma k'$})}{2m_\Lambda}\
\chi^c_{\Lambda \ell '} \\ \chi^c_{\Lambda \ell '}
\end{array}\right),\nn \\ && \bar \psi^c_{\Lambda \ell}(-p_2) \simeq
-i\sqrt{2m_\Lambda} \left(\chi^{c+}_{\Lambda \ell}
\frac{- (\mbox{\boldmath$\sigma
k$})}{2m_\Lambda} \,,-\chi^{c+}_{\Lambda \ell}\right), \nn
\end{eqnarray}
where
$\varphi_{Nj}$ and $\chi^c_{\Lambda\ell}$ are two-component spinors.
For the waves with $J=0,1$ we have
\begin{eqnarray}
&&\hspace{-7mm}L=0, J=0:\quad
\left(\bar\psi_N (p'_1)
\hat Q^{000}(k')\psi^c_\Lambda(-p'_2)\right)
\nn \\ &&\times
\left(\bar\psi^c_\Lambda(-p_2)
\hat Q^{000}(k)\psi_N(p_1)\right)
A^{(0,00,0)}_{N\Lambda\to N\Lambda}(s) \simeq \nn \\
&&
\sqrt{4m_N m_\Lambda}
\left( \varphi^+_{Nj'} \chi^c_{\Lambda \ell '}\right)
\left(\chi^{c+}_{\Lambda \ell }\varphi_{Nj}\right)
\sqrt{4m_N m_\Lambda}\,\nn  \\
&&\times A^{(0,00,0)}_{N\Lambda\to N\Lambda}(s), \nn \\
&&\hspace{-7mm}L=0, J=1: \quad
\left(\bar\psi_N (p'_1)
\hat Q^{101}_{\mu}(k')\psi^c_\Lambda(-p'_2)\right)
\nn \\ &&\times
\left(\bar\psi^c_\Lambda(-p_2)
\hat Q^{101}_{\mu}(k)\psi_N(p_1)\right)
A^{(1,00,1)}_{N\Lambda\to N\Lambda}(s) \simeq \nn \\
&&
i\sqrt{4m_N m_\Lambda}\left( \varphi^+_{Nj'}
\mbox{\boldmath$\sigma $}\chi^c_{\Lambda
\ell '}\right)
\left(\chi^{c+}_{\Lambda \ell}
\mbox{\boldmath$\sigma $}\varphi_{j}\right)
i \sqrt{4m_N m_\Lambda}\,
\nn \\ &&\times
A^{(1,00,1)}_{N\Lambda\to N\Lambda}(s)\, .
\label{TB29}
\end{eqnarray}
Let us consider bispinors with real components. For nucleons, we write
\be \label{TB30}
\varphi_{Nj}=
\left(\begin{array}{c}\varphi_{\uparrow}(Nj) \\ \varphi_{\downarrow}(Nj)
\end{array} \right), \;
\varphi^+_{Nj}=
\left(\varphi_{\uparrow}(Nj), \varphi_{\downarrow}(Nj)
\right).
\ee
For the $\Lambda$, we determine the bispinor to be given by
\be \label{TB31}
\chi^c_{\Lambda\ell}=i\sigma_2
\left(\begin{array}{c}\varphi_{\uparrow}(\Lambda\ell)
\\ \varphi_{\downarrow}(\Lambda\ell) \end{array} \right)=
\left(\begin{array}{c}\varphi_{\downarrow}(\Lambda\ell)
\\ -\varphi_{\uparrow}(\Lambda\ell) \end{array} \right)
\ee
Within this definition, we can re-write (\ref{TB29})  in terms
of the traditional technique which uses the Clebsch-Gordan coefficients.
For $J=0$, we have
\be \label{TB32}
&&\left(\chi^{c+}_{\Lambda \ell }\frac{I}{\sqrt 2}\varphi_{Nj}\right) =
\left( \varphi^+_{Nj}\frac{I}{\sqrt 2} \chi^c_{\Lambda \ell }\right)
= \nn \\ &&
 \frac{1}{\sqrt 2} \left(\varphi_{\uparrow}(Nj)
\varphi_{\downarrow}(\Lambda\ell)-
       \varphi_{\downarrow}(Nj)\varphi_{\uparrow}(\Lambda\ell)\right)=\nn \\
&=&\sum\limits_{\alpha}C^{00}_{1/2\alpha\ ,\ 1/2-\alpha}\,
\varphi_{\alpha}(Nj)\varphi_{-\alpha}(\Lambda\ell),
\ee
and for $J=1,J_3=0$,
\be \label{TB33}
&& \left(\chi^{c+}_{\Lambda \ell }\frac{\sigma_3}{\sqrt
2}\varphi_{Nj}\right) = \left( \varphi^+_{Nj}\frac{\sigma_3}{\sqrt 2}
\chi^c_{\Lambda \ell }\right) = \nn \\ &&
 \frac{1}{\sqrt 2}
\left(\varphi_{\uparrow}(Nj) \varphi_{\downarrow}(\Lambda\ell)+
      \varphi_{\downarrow}(Nj)\varphi_{\uparrow}(\Lambda\ell)\right)=\nn \\
&&\sum\limits_{\alpha}C^{10}_{1/2\alpha\ ,\ 1/2-\alpha}\,
\varphi_{\alpha}(Nj)\varphi_{-\alpha}(\Lambda\ell).
\ee

{\bf (ii) The $D$-wave component in the  operator {\boldmath
$\gamma^\perp_\mu$}.}

Equations (\ref{TB28}) and (\ref{TB29}) allow one to see easily the
existence of the $D$-wave admixture in the operator $\gamma^\perp_\mu$.
By using the operator $\hat Q^{101}_{\mu}(k) = \gamma^\perp_\mu$ in
(\ref{TB29}), one has the following next-to-leading term in the
($J=1$)-wave:
\be  \label{TB34}
&&\hspace{-9mm}
-\sqrt{4m_N m_\Lambda}\left( \varphi^+_{Nj'}
\frac{(\mbox{\boldmath$\sigma k'$})}{2m_N}
\mbox{\boldmath$\sigma $}\frac{(\mbox{\boldmath$\sigma
k'$})}{2m_\Lambda} \chi^c_{\Lambda \ell '}\right)
\nn \\ &&
\hspace{-9mm}\times
\left(\chi^{c+}_{\Lambda \ell}
\frac{(\mbox{\boldmath$\sigma
k$})}{2m_\Lambda} \mbox{\boldmath$\sigma
$}\frac{(\mbox{\boldmath$\sigma k$})}{2m_N}
\varphi_{Nj}\right)
\sqrt{4m_N m_\Lambda}\, A^{(1,00,1)}_{N\Lambda\to N\Lambda}(s) .
\end{eqnarray}
The spin operators in (\ref{TB34}) can be presented as \be \label{TB35}
\frac{(\mbox{\boldmath$\sigma k$})}{2m_\Lambda} \mbox{\boldmath$\sigma
$}\frac{(\mbox{\boldmath$\sigma k$})}{2m_N} \simeq \frac{\mbox{\boldmath$
k$}(\mbox{\boldmath$\sigma k$})}{2m_\Lambda m_N} + \mbox{\boldmath$\sigma $} \,
O\left(\frac{\mbox{\boldmath$ k$}^2}{m_\Lambda m_N} \right), \ee where the
first term in the r.-h. side refers to the $D$-wave, while the second one gives
the correction to the $S$-wave term.  In the operator $\Gamma_\alpha(k_\perp)$,
the $D$-wave admixture is canceled due to the second term: $ -[4s
k_{\perp\alpha} (k_{\perp}\gamma)]/$ $
[(m_N+m_\Lambda)(\sqrt{s}+m_{N}+m_{\Lambda} )(s-(m_N-m_\Lambda)^2)]$

\section* {Appendix 3. Useful relations for \boldmath{$Z^\alpha_{\mu_1\ldots\mu_n}$}
and \boldmath{$X^{(n-1)}_{\nu_2\ldots\nu_n}$} }

In this appendix, we list a few useful expressions.
\be
&&Z^\alpha_{\mu_1\ldots\mu_n}\;=\;X^{(n-1)}_{\nu_2\ldots\nu_n}
O^{\alpha\nu_2\ldots\nu_n}_{\mu_1\ldots\mu_n} \frac{2n-1}{n}\ ,
\label{App_1} \\
&&Z^\alpha_{\mu_1\ldots\mu_n}(q)
(-1)^n O^{\mu_1\ldots\mu_n}_{\nu_1\ldots\nu_n}
Z^\beta_{\nu_1\ldots\nu_n}(k)=
\frac{\alpha_n}{n^2}(-1)^n
\nn \\ &&\times
\left(\sqrt{k^2_\perp}\sqrt{q^2_\perp}\right)^{n-1}
\left[g^\perp_{\alpha\beta}P'_n
-\left(
 \frac{q^\perp_\alpha q^\perp_\beta}{q^2_\perp}
+\frac{k^\perp_\alpha k^\perp_\beta}{k^2_\perp}
\right)P''_{n-1} \right.
\nn \\
&&\left. +\frac{q^\perp_\alpha
k^\perp_\beta}{\sqrt{k^2_\perp}\sqrt{q^2_\perp}}
\left(P''_{n-2}-2P'_{n-1}\right)
+\frac{k^\perp_\alpha
q^\perp_\beta}{\sqrt{k^2_\perp}\sqrt{q^2_\perp}} P''_n\right],
\\
&&X_{\alpha\mu_1\ldots\mu_n}(q)
(-1)^n O^{\mu_1\ldots\mu_n}_{\nu_1\ldots\nu_n}
X_{\beta\nu_1\ldots\nu_n}(k)=
\frac{\alpha_n}{(n+1)^2}(-1)^n
\nn \\
&&\times
\left(\sqrt{k^2_\perp}\sqrt{q^2_\perp}\right)^{n+1}
\left[g^\perp_{\alpha\beta}P'_{n+1}
\!-\!\left(
 \frac{q^\perp_\alpha q^\perp_\beta}{q^2_\perp}
+\frac{k^\perp_\alpha k^\perp_\beta}{k^2_\perp}
\right)P''_{n+1} \right.
\nn \\
&&\left. +\frac{q^\perp_\alpha
k^\perp_\beta}{\sqrt{k^2_\perp}\sqrt{q^2_\perp}}
\left(P''_{n+2}-2P'_{n+1}\right)
+\frac{k^\perp_\alpha q^\perp_\beta}{\sqrt{k^2_\perp}\sqrt{q^2_\perp}}
P''_n\right],
\\
&&Z^\alpha_{\mu_1\ldots\mu_n}(q)
(-1)^n O^{\mu_1\ldots\mu_n}_{\nu_1\ldots\nu_n}
X_{\beta\nu_1\ldots\nu_n}(k)= \frac{\alpha_{n-1}}{n(n+1)}(-1)^n
\nn \\
&&
\times(-k^2_\perp)\left(\sqrt{k^2_\perp}\sqrt{q^2_\perp}\right)^{n+1}
\left[g^\perp_{\alpha\beta}P'_n
-\frac{q^\perp_\alpha q^\perp_\beta}{q^2_\perp}P''_{n-1}
\right.
\nn \\
&&\left.
-\frac{k^\perp_\alpha k^\perp_\beta}{k^2_\perp}P''_{n+1}
+\frac{q^\perp_\alpha k^\perp_\beta}{\sqrt{k^2_\perp}\sqrt{q^2_\perp}}
P''_n
+\frac{k^\perp_\alpha q^\perp_\beta}{\sqrt{k^2_\perp}\sqrt{q^2_\perp}}
P''_n\right].
 \ee

We now consider some further expressions used in the
one-loop diagram calculations. In our case, the
operators are constructed  of
$X^{(n+1)}_{\alpha\mu_1\ldots\mu_n}$ and
$Z^\beta_{\mu_1\ldots\mu_n}$, where $\alpha$ and $\beta$ indices to
be convoluted with tensors.
 Let us start with the loop diagram with a $Z$-operator:
 \be
\!\int\!\frac{d\Omega}{4\pi}
Z^\alpha_{\mu_1\ldots\mu_n}(k^\perp)
T_{\alpha\beta}
Z^\beta_{\nu_1\ldots\nu_n}(k^\perp)\!=\!\lambda
O^{\mu_1\ldots\mu_n}_{\nu_1\ldots\nu_n}(-1)^n . \quad\;\;
 \ee
For  different tensors $T_{\alpha\beta}$, one has the following
$\lambda$'s:
 \be
&T_{\alpha\beta}=g_{\alpha\beta},\qquad\qquad &
\lambda=-\frac{\alpha_n}{n}|\vec k|^{2n-2}\;,  \label{app_loop1}
\\
&T_{\alpha\beta}=k^\perp_\alpha k^\perp_\beta,\qquad\qquad &
\lambda=\frac{\alpha_n}{2n+1}|\vec k|^{2n}\ .     \label{app_loop2}
 \ee
Equation (\ref{app_loop1})
can be easily obtained using eqs. (\ref{App_1}) and (\ref{x-prod}),
while Eq.(\ref{app_loop2}) can be obtained using eqs.
(\ref{z}) and (\ref{x-prod}).
For the $X$ operators, one has
 \be
\!\int\frac{d\Omega}{4\pi}
X^{(n+1)}_{\alpha\mu_1\ldots\mu_n}(k^\perp)
T_{\alpha\beta}
X^{(n+1)}_{\beta\nu_1\ldots\nu_n}(k^\perp)\!=\!\lambda
O^{\mu_1\ldots\mu_n}_{\nu_1\ldots\nu_n}(-1)^n, \nn \\
 \label{app_loop3}\ee
where
 \be
&T_{\alpha\beta}=g_{\alpha\beta},\qquad\qquad &
\lambda=-\frac{\alpha_n}{n+1}|\vec k|^{2n+2},
\nn \\
&T_{\alpha\beta}=k^\perp_\alpha k^\perp_\beta,\qquad\qquad &
\lambda=\frac{\alpha_n}{2n+1}|\vec k|^{2n+4}.  \label{app_loop4}
 \ee
To derive Eq. (\ref{app_loop3}), the properties
\be
O^{\alpha\mu_1\ldots\mu_n}_{\alpha\nu_1\ldots\nu_n} = \frac{2n+3}{2n+1}
O^{\mu_1\ldots\mu_n}_{\nu_1\ldots\nu_n}
\ee
of the projection operator and Eq. (\ref{ceq}) are used.
The interference term between $X$ and $Z$ operators is given by
 \be
\!\!\!\int\!\!\frac{d\Omega}{4\pi}
X^{(n+1)}_{\alpha\mu_1\ldots\mu_n}(k^\perp)
T_{\alpha\beta}
Z^{\beta}_{\nu_1\ldots\nu_n}(k^\perp)\!=\!\lambda
O^{\mu_1\ldots\mu_n}_{\nu_1\ldots\nu_n}(-1)^n ,\nn \\
 \ee
with
 \be
&T_{\alpha\beta}=g_{\alpha\beta},\qquad\quad &
\lambda=0,             \label{app_loop5}
\nn \\
&T_{\alpha\beta}=k^\perp_\alpha k^\perp_\beta,\qquad\quad &
\lambda=-\frac{\alpha_n}{2n+1}|\vec k|^{2n+2}. \label{app_loop6}
 \ee
 Eq. (\ref{app_loop5}) is calculated using Eq. (\ref{App_1}) and
the  orthogonality properties (\ref{ort-x}) of the $X$ operators.


\section*{Appendix 4. \boldmath{$N\Delta$} one-loop diagrams}

The calculation of the one-loop diagram for different vertex operators
is an important step in the construction of the unitary $N\Delta$ amplitude.
Consider the loop diagram for  $S=1$ and derive all expressions
in case of different particle masses ($m_1$ is mass of $\Delta$ and $m_2$ is nucleon
mass).

Let us start with $^3L_J$ $(J=L-1)$ states.
Using the  expression
 \be
&&Sp\Big[i\gamma_5(m_1+\hat k_1)
\Big(g^{\perp k_1}_{\alpha\beta} -
 \frac{\gamma^{\perp k_1}_\alpha
\gamma^{\perp k_1}_\beta}{3}\Big)
i\gamma_5(m_2- \hat k_2)\Big]
\nn \\ &&=
-\frac43 \Big(g_{\alpha\beta} - \frac {k^\perp_\alpha k^\perp_\beta}{m^2_1}\Big)
(s-\delta^2),
\label{NDloop_1}
 \ee
where $\delta = m_1 -m_2$, the one-loop diagram for the operator
(\ref{operW1}) is given by
 \be
&&\int\frac{d\Omega}{4\pi}Sp\Big[V^{(1)\alpha}_{\mu_1\ldots\mu_n}(m_1+\hat k_1)
\Big(g^{\perp k_1}_{\alpha\beta} -
 \frac{\gamma^{\perp k_1}_\alpha
\gamma^{\perp k_1}_\beta}{3}\Big)
\nn \\ &&
\qquad\qquad\times V^{(1)\beta}_{\nu_1\ldots\nu_n}(m_2-\hat k_2)\Big]=
\nn \\
&& \frac43 (s-\delta^2)\frac{\alpha_n}{n+1}
\Big(1+\frac{|\vec k|^2 (n+1)}{m^2_1(2n+1)}\Big)
|\vec k|^{2n+2}
\nn \\
&&\times O^{\mu_1\ldots\mu_n}_{\nu_1\ldots\nu_n}(-1)^{n}.
\label{loop_w1}
 \ee
Here, eqs. (\ref{app_loop3}) and (\ref{app_loop4}) were used.

For $^3L_J$ $(J=L+1)$ states, one has
 \be
&&\int\frac{d\Omega}{4\pi}Sp\Big[V^{(2)\alpha}_{\mu_1\ldots\mu_n}(m_1+\hat k_1)
\Big(g^{\perp k_1}_{\alpha\beta} -
 \frac{\gamma^{\perp k_1}_\alpha
\gamma^{\perp k_1}_\beta}{3}\Big)
\nn \\ &&
\qquad\qquad\times V^{(2)\beta}_{\nu_1\ldots\nu_n}(m_2-\hat k_2)] = \nn
\\
&& \frac43 (s-\delta^2)\frac{\alpha_{n-1}}{2n-1}\Big(1+\frac{|\vec k|^2 n}{m^2_1
(2n+1)}\Big)|\vec k|^{2n-2}
\nn \\
&&\times O^{\mu_1\ldots\mu_n}_{\nu_1\ldots\nu_n}(-1)^{n}.
\label{loop_w2}
 \ee
Direct calculations also show that transition loop diagrams between
$^3L_J$ $(J=L-1)$ and $^3L_J$ $(J=L+1)$ states
are equal to
 \be
&&\int\frac{d\Omega}{4\pi}Sp\Big[V^{(1)\alpha}_{\mu_1\ldots\mu_n}(m_1+\hat k_1)
\Big(g^{\perp k_1}_{\alpha\beta} -
\frac{\gamma^{\perp k_1}_\alpha
\gamma^{\perp k_1}_\beta}{3}\Big)
\nn \\ &&
\qquad\qquad\times V^{(2)\beta}_{\nu_1\ldots\nu_n}(m_2-\hat k_2)\Big] =
\nn \\ &&
 \frac43 (s-\delta^2)\frac{\alpha_{n-1}}{2n+1}
\frac{|\vec k|^{2n+2}}{m^2_1}
O^{\mu_1\ldots\mu_n}_{\nu_1\ldots\nu_n}(-1)^{n+1}.
\label{loop_w1w2}
 \ee

One can also introduce the pure spin operator in a way that the
transition loop diagram is equal to zero. Then
eqs.(\ref{operW1}-\ref{operW3}) can be rewritten in the following way:
 \be
W^{(i)}_{\mu_1\ldots\mu_n} =\bar \psi_\alpha(k_1) \Gamma^{3/2}_{\alpha\beta}
V^{(i)\beta}_{\mu_1\ldots\mu_n} u(-k_2)\;,\;\; i=1,2\ ,\qquad
 \ee
where
\be
\Gamma^{3/2}_{\alpha\beta} = g_{\alpha\beta}+
\frac{4s k^\perp_\alpha k^\perp_\beta}{(s+M\delta)(\sqrt{s}+M)(\sqrt{s}+\delta)}.
\ee
Then, it is easy to find that
 \be
&&Sp\Big[i\gamma_5(m_1+\hat k_1) \Gamma^{3/2}_{\alpha\alpha'}
\Big(g^{\perp k_1}_{\alpha'\beta'} -
 \frac{\gamma^{\perp k_1}_{\alpha'}
\gamma^{\perp k_1}_{\beta'}}{3}\Big)\Gamma^{3/2}_{\beta\beta'}
\nn \\&&\quad\times
i\gamma_5(m_2- \hat k_2)\Big]
=
-\frac43 g_{\alpha\beta}(s-\delta^2)\ .
\label{NDloop_11}
 \ee
Thus, the transition loop diagram vanishes identically.

For $^3L_J$ $(J=L)$ states, one has:
 \be
&&\int\frac{d\Omega}{4\pi}Sp\Big[V^{(3)\alpha}_{\mu_1\ldots\mu_n}(m_1+\hat k_1)
\Big(g^{\perp k_1}_{\alpha\beta} -
 \frac{\gamma^{\perp k_1}_\alpha
\gamma^{\perp k_1}_\beta}{3}\Big)
\nn \\ &&
\qquad\qquad\times V^{(3)\beta}_{\nu_1\ldots\nu_n}(m_2-\hat k_2)\Big] =
\nn \\
&& \frac43 (s-\delta^2)s \alpha_{n-1} \frac{n+1}{4n^2-1}
|\vec k|^{2n} O^{\mu_1\ldots\mu_n}_{\nu_1\ldots\nu_n}(-1)^{n}.
\label{loop_w3}
 \ee
To calculate loop diagrams with $S=2$, the following expression is used:
 \be
&&O^{\alpha_1\alpha_2}_{\mu_1\mu_2}
Sp\Big[\gamma_{\mu_1}(m_1+\hat k_1)
\Big(g^{\perp k_1}_{\alpha\beta} -
 \frac{\gamma^{\perp k_1}_\alpha
\gamma^{\perp k_1}_\beta}{3}\Big)
\nn \\
 &&\qquad\qquad \times
\gamma_{\nu_2}(m_2- \hat k_2)\Big] O^{\nu_1\nu_2}_{\beta_1\beta_2}=
\nn \\
 &&
a_1 O^{\alpha_1\alpha_2}_{\beta_1\beta_2} +
a_2 Z^{\xi}_{\alpha_1\alpha_2}Z^{\xi}_{\beta_1\beta_2} +
 a_3 X^{(2)}_{\alpha_1\alpha_2}X^{(2)}_{\beta_1\beta_2},
 \label{NDloop_s21}
 \ee
where
\be
&&a_1 = 2(s-\delta^2)\;, \;\; a_2 = \frac{32\delta}{9m_1} - \frac{16}{27m_1^2}(s-(m_1 +
m_2)^2)\;,\nn \\
&&a_3 = - \frac{64}{27m_1^2}\ .
 \ee
For $^5L_J$ $(J=L+2)$, the operator one-loop diagram is equal to
 \be
&&\int\frac{d\Omega}{4\pi}Sp\Big[V^{(4)\alpha}_{\mu_1\ldots\mu_n}(m_1+\hat k_1)
\Big(g^{\perp k_1}_{\alpha\beta} -
 \frac{\gamma^{\perp k_1}_\alpha
\gamma^{\perp k_1}_\beta}{3}\Big)
\nn \\ &&
\qquad\qquad\quad \times
V^{(4)\beta}_{\nu_1\ldots\nu_n}(m_2-\hat k_2)\Big] =
\nn \\
&& \frac{\alpha_{n-2}}{2n-3} |\vec k|^{2n-4} (-1)^{n}
O^{\mu_1\ldots\mu_n}_{\nu_1\ldots\nu_n}
\nn \\
&&\times\Big( a_1 + \frac94 \frac{n-1}{2n-1} (-a_2 |\vec k|^2 + a_3 \frac{n}{2n+1}|\vec k|^4)
 \Big).
\label{loop_w4}
 \ee
For $^5L_J$ $(J=L-2)$, the one-loop operator is given by
 \be
&&\int\frac{d\Omega}{4\pi}Sp\Big[V^{(5)\alpha}_{\mu_1\ldots\mu_n}(m_1+\hat k_1)
\Big(g^{\perp k_1}_{\alpha\beta} -
 \frac{\gamma^{\perp k_1}_\alpha
\gamma^{\perp k_1}_\beta}{3}\Big)
\nn \\ &&\qquad\qquad\times
V^{(5)\beta}_{\nu_1\ldots\nu_n}(m_2-\hat k_2)\Big] =
\nn \\
&& \alpha_{n} |\vec k|^{2n+4} (-1)^{n}
O^{\mu_1\ldots\mu_n}_{\nu_1\ldots\nu_n}
\nn \\
&&\times\Big( \frac{(2n+3)a_1}{(n+1)(n+2)} + \frac94 \Big(-\frac{a_2 |\vec k|^2}{n+1} +
\frac{a_3 |\vec k|^4}{2n+1}\Big)\Big), \qquad \quad
\label{loop_w5}
 \ee
while for $^5L_J$ $(J=L)$, the operator is written as
 \be
&&\int\frac{d\Omega}{4\pi}Sp\Big[V^{(6)\alpha}_{\mu_1\ldots\mu_n}(m_1+\hat k_1)
\Big(g^{\perp k_1}_{\alpha\beta} -
 \frac{\gamma^{\perp k_1}_\alpha
\gamma^{\perp k_1}_\beta}{3}\Big)
\nn \\ &&\qquad\qquad\times
V^{(6)\beta}_{\nu_1\ldots\nu_n}(m_2-\hat k_2)\Big] =
\nn \\
&& \frac{\alpha_{n-1}}{2n(2n+1)} |\vec k|^{2n} (-1)^{n}
O^{\mu_1\ldots\mu_n}_{\nu_1\ldots\nu_n}
\nn \\
&&\times\Big[ \frac{(2n+3)(n+1)a_1}{3n} - \frac98 |\vec k|^2 a_2
\Big(\frac{2n+5}{9}+\frac{2n+1}{n(2n-1)}\Big)
\nn \\
 &&\quad+\frac{a_3 |\vec k|^4 (n+1)^2}{2(2n-1)}
 \Big].
\label{loop_w6}
 \ee
The one-loop transition diagram between $^5L_J$ $(J=L+2)$ and $^5L_J$
$(J=L-2)$ states can be expressed as
 \be
&&\int\frac{d\Omega}{4\pi}Sp\Big[V^{(4)\alpha}_{\mu_1\ldots\mu_n}(m_1+\hat k_1)
\Big(g^{\perp k_1}_{\alpha\beta} -
 \frac{\gamma^{\perp k_1}_\alpha
\gamma^{\perp k_1}_\beta}{3}\Big)
 \\ &&\times
V^{(5)\beta}_{\nu_1\ldots\nu_n}(m_2-\hat k_2)\Big]
=\frac 94 \frac{\alpha_{n-2}}{2n+1}  a_3 |\vec k|^{2n+4} (-1)^{n}
O^{\mu_1\ldots\mu_n}_{\nu_1\ldots\nu_n},     \nonumber
\label{loop_w45}
 \ee
and the one-loop transition  diagram between $^5L_J$ $(J=L+2)$ and
$^5L_J$ $(J=L)$ states as
 \be
&&\int\frac{d\Omega}{4\pi}Sp\Big[V^{(4)\alpha}_{\mu_1\ldots\mu_n}(m_1+\hat k_1)
\Big(g^{\perp k_1}_{\alpha\beta} -
 \frac{\gamma^{\perp k_1}_\alpha
\gamma^{\perp k_1}_\beta}{3}\Big)
\nn \\ &&\qquad\qquad\times
V^{(6)\beta}_{\nu_1\ldots\nu_n}(m_2-\hat k_2)\Big] =
\nn \\
&&\frac{3\alpha_{n-2}(n+1)}{8(2n+1)(2n-1)} |\vec k|^{2n} (-1)^{n}
O^{\mu_1\ldots\mu_n}_{\nu_1\ldots\nu_n}
\nn \\ &&\qquad\qquad
\times\Big( \frac{2n+3}{n}a_2 - 2 |\vec k|^{2}a_3 \Big)\ .
\label{loop_w45_1}
 \ee
For the one-loop transition diagram between $^5L_J$ $(J=L-2)$ and
$^5L_J$ $(J=L)$ we get
 \be
&&\int\frac{d\Omega}{4\pi}Sp\Big[V^{(5)\alpha}_{\mu_1\ldots\mu_n}(m_1+\hat k_1)
\Big(g^{\perp k_1}_{\alpha\beta} -
\frac{\gamma^{\perp k_1}_\alpha
\gamma^{\perp k_1}_\beta}{3}\Big)
\nn \\ &&\qquad\qquad\times
V^{(6)\beta}_{\nu_1\ldots\nu_n}(m_2-\hat k_2)\Big] =
\\
&&\frac{3\alpha_{n}}{8(2n+1)} |\vec k|^{2n+4} (-1)^{n}
O^{\mu_1\ldots\mu_n}_{\nu_1\ldots\nu_n}
\Big( a_2 - 2 |\vec k|^{2}a_3 \frac{n+1}{2n-1}\Big)\ , \nn
\label{loop_w56}
 \ee
and, for $^5L_J$ $(J=L-1)$, we find
to
 \be
&&\int\frac{d\Omega}{4\pi}Sp\Big[V^{(7)\alpha}_{\mu_1\ldots\mu_n}(m_1+\hat k_1)
Big(g^{\perp k_1}_{\alpha\beta} -
\frac{\gamma^{\perp k_1}_\alpha
\gamma^{\perp k_1}_\beta}{3}\Big)
\nn \\ &&\qquad\qquad\times
V^{(7)\beta}_{\nu_1\ldots\nu_n}(m_2-\hat k_2)\Big] =
\nn \\
&& \frac{s\alpha_{n-1}}{2(2n+1)} |\vec k|^{2n+2} (-1)^{n}
O^{\mu_1\ldots\mu_n}_{\nu_1\ldots\nu_n}
\nn \\
&&\times\Big( \frac{a_1(n+1)(2n^2+n-2)}{n^2(2n-1)} -
\frac98 |\vec k|^2 a_2 \frac{n+1}{2n-1} \Big)\ .
\label{loop_w7}
 \ee
Finally, the operator one-loop diagram For $^5L_J$ $(J=L+1)$ is equal to
 \be
&&\int\frac{d\Omega}{4\pi}Sp\Big[V^{(8)\alpha}_{\mu_1\ldots\mu_n}(m_1+\hat k_1)
\Big(g^{\perp k_1}_{\alpha\beta} -
 \frac{\gamma^{\perp k_1}_\alpha
\gamma^{\perp k_1}_\beta}{3}\Big)
\nn \\ &&\qquad\qquad\times
V^{(8)\beta}_{\nu_1\ldots\nu_n}(m_2-\hat k_2)\Big] =
\nn \\
&& \frac{s\alpha_{n-2}(n+1)}{2(2n-1)(2n-3))} |\vec k|^{2n-2} (-1)^{n}
O^{\mu_1\ldots\mu_n}_{\nu_1\ldots\nu_n}
\nn \\
&&\times\Big( a_1 - \frac98 |\vec k|^2 a_2 \frac{n-1}{2n+1} \Big) ,
\label{loop_w8}
 \ee
and the one-loop transition diagram between $^5L_J$ $(J=L-1)$ and
$^5L_J$ $(J=L+1)$ can be written as
 \be
&&\int\frac{d\Omega}{4\pi}Sp\Big[V^{(7)\alpha}_{\mu_1\ldots\mu_n}(m_1+\hat k_1)
\Big(g^{\perp k_1}_{\alpha\beta} -
 \frac{\gamma^{\perp k_1}_\alpha
\gamma^{\perp k_1}_\beta}{3}\Big)
\nn \\ &&\qquad\qquad\times
V^{(8)\beta}_{\nu_1\ldots\nu_n}(m_2-\hat k_2)\Big] =
\\
&& \frac{s\alpha_{n-2}}{4n^2\!-\!1} |\vec k|^{2n} (-1)^{n}
O^{\mu_1\ldots\mu_n}_{\nu_1\ldots\nu_n}
\Big( \frac{n\!+\!1}{n} a_1 +
\frac{9}{16}|\vec k|^2 a_2 (n\!+\!1) \Big)\ .\nn
\label{loop_w78}
 \ee

\newpage

\section*{Appendix 5. Amplitude of the triangle diagram}

\begin{figure}[h]
\vspace{-5mm}
\centerline{
\epsfig{file=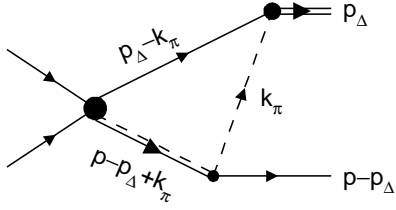,width=0.30\textwidth,clip=on}}
\caption{\label{fig_triangle1}
Triangle diagram.}
\end{figure}
In the last two appendices, we give results on triangle diagrams
in numerical form. First, we calculate the triangle-diagram integral
which enters  Eq. (\ref{D20}):
\be
\label{tr1}
&&A_{\rm triangle}^{\rm
spinless}(W^2,s)= \int
\frac{d^4k_\pi}{i(2\pi)^4}\frac{1}{m^2_\pi-k^2_\pi-i0}
 \\
&& \times
 \frac{1}{m^2_\Delta\!-\!(p\!-\!p_{\Delta}\!+\!k_\pi)^2-im_\Delta \Gamma_\Delta}
\frac{1}{m^2_N-(p_\Delta-k_\pi)^2-i0} \, .
 \nn
\ee
\begin{figure}[pt]
\leftline{\epsfig{file=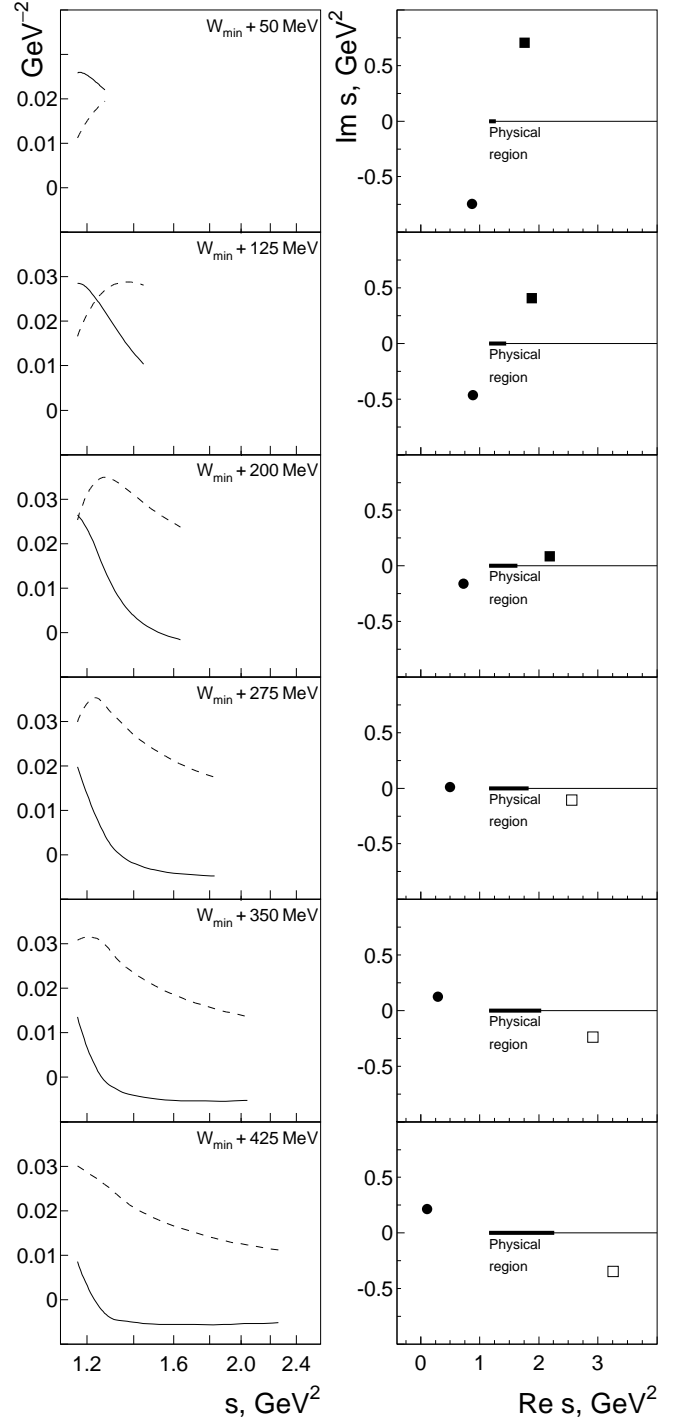,width=0.49\textwidth,clip=on}}
\caption{Triangle diagram amplitude.
In the left columns, real and imaginary parts of the amplitude are shown
by solid and dashed curves, correspondingly. Initial energy, $W$,
is shown on the top of each panel.
In the right columns, singularity positions, $s^{(tr)}_{(\pm)}$, Eq.(\ref{J_sing}), are
shown on the 2nd sheet of the complex-$s$ plane. When $s^{(tr)}_{(+)}$
dives to the 3rd sheet, its position is shown by open square.}
\label{fig_triangle_02}
 \end{figure}
 Notations of the momenta are illustrated by Fig. \ref{fig_triangle1}.
Here,
 \be \label{tr2}
 p=p_1+p_2,\quad p^2=W^2,\quad  p^2_\Delta =s\, .
\ee
The physical region is located in the interval:
\be \label{tr3}
 (m_N+m_\pi)^2\le &s&\le(W-m_N)^2\, .
 \ee

The triangle-diagram amplitude $\,$ $A_{\rm triangle}^{\rm spinless}(W^2,s)$
determined by (\ref{tr1}) is shown in
the physical region (\ref{tr3}) in Fig. \ref{fig_triangle_02} (left column).
In the right
column, there are positions  of the logarithmic singularities on the
second sheet of the complex-$s$ plane. Physical region of the reaction
is also shown (thick solid line): it is located on the lower edge of
the cut related to the threshold singularity (thin solid line). The
positions of logarithmic singularities read:
 \be
&& s^{(tr)}_{(\pm)}=m^2_\pi\!+\!m^2_N
\!+\!\frac{(W^2-M^2_\Delta-m^2_N)
       (M^2_\Delta+m^2_\pi-m^2_N)}
{2M^2_\Delta}  \nn \\
&&\pm
\Big[(m^2_\pi-(M_\Delta-m_N)^2)
 (m^2_\pi-(M_\Delta+m_N)^2) \nn \\
&&\times
 (W^2-(M_\Delta-m_N)^2)
 (W^2-(M_\Delta+m_N)^2)\Big]^{1/2}. \label{J_sing}
 \ee
Here $M^2_\Delta=m^2_\Delta-im_\Delta\Gamma_\Delta$.

In the left column of Fig. \ref{fig_triangle_02}, the real and
imaginary parts of the amplitude (\ref{tr1}) at different total
energies $W$ are shown by solid and dashed curves, respectively. In the
right column, one sees the singularity positions, $s^{(tr)}_{(-)}$
(black circles) and $s^{(tr)}_{(+)}$ (black squares). When
$s^{(tr)}_{(+)}$ dives into the third sheet, its position is shown as
an open square.


\section*{Appendix 6. Amplitude of the box diagram}
Here, we calculate the box-diagram integral which enters Eq.
(\ref{2-9}), the notations of momenta  are given in Fig. \ref{fig_box}.

\begin{figure}[h]
\vspace{-3mm}
\centerline{
\epsfig{file=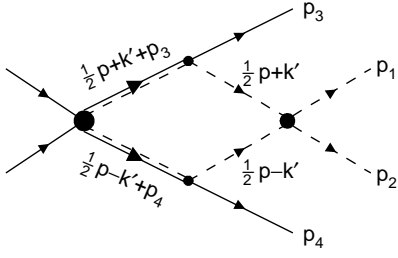,width=0.30\textwidth}}
\caption{\label{fig_box}
Box diagram.}
\vspace{-3mm}
\end{figure}
\be
&&A_{\rm box}^{\rm spinless}(W^2,s_3,s_4,s_{12})=
\nn \\
&&\int\frac{d^4k'}{i(2\pi)^4}
\frac{1}{(m^2_{\Delta}-(\frac 12 p+k'+p_3)^2-im_{\Delta}\Gamma_{\Delta})}
\nn \\
&&\times
\frac{1}{(m^2_\Delta-(\frac 12 p-k'+p_4)^2-im_\Delta\Gamma_\Delta)}
\nn \\
&&\times
\frac{1}{(m^2_{\pi}-(\frac 12 p+k')^2-i0)(m^2_{\pi}-(\frac 12 p-k')^2-i0)}.
\qquad
\ee
Remind that $s_3=(p-p_3)^2$, $s_4=(p-p_4)^2$, $s_{12}=(p_1+p_2)^2$, $W^2=p^2$.

In Fig. \ref{fig_box_02}, we show the results of our  calculation
of $A_{\rm box}^{\rm spinless}(W^2,s_3,s_4,s_{12})$ as a
function of pion-pion energy squared $s_{12}$ at different total
energies $W$, under the following constraint on $s_3$ and $s_4$:
\be
s_3=s_4=\sqrt{s_{12}}\, W+m_N^2\, .
\label{constraint}
\ee

This constraint corresponds to the following kinematics in the c.m.
system:
 \be
 p&=&(W;0;0;0), \nn \\
 p_1&=&(\sqrt{m^2_\pi+p^2_{1z}};0;0;p_{1z}),\nn \\
 p_2&=&(\sqrt{m^2_\pi+p^2_{1z}};0;0;-p_{1z}),\nn \\
 p_3&=&(\sqrt{m^2_N+p^2_{3z}};0;0;p_{3z}),\nn \\
 p_4&=&(\sqrt{m^2_N+p^2_{3z}};0;0;-p_{3z}), \nn \\
&&\sqrt{m^2_\pi+p^2_{1z}}+ \sqrt{m^2_N+p^2_{3z}} =W/2
\ee
The positions of the box-diagram singularities are given by the formula:
 \be
\label{a6box}
&&s_{12}^{\rm box}=2m^2_\pi+\frac 1{2W^2}(s_3-m^2_N)(s_4-m^2_N)\\
&+&
\frac{(2W^2M^2_\Delta\!-\!W^2(s_3-m^2_N))
      (2W^2M^2_\Delta\!-\!W^2(s_4-m^2_N))}
{2W^2((W^2-2M^2_\Delta)^2-4M^4_\Delta)} \nn \\
&-&\left[\left(\frac{(s_3\!-\!m^2_N)^2}{2W^2}
\!-\!2m^2_\pi
\!-\!\frac{(2W^2M^2_\Delta\!-\!W^2(s_3\!-\!m^2_N))^2}
      {2W^2((W^2\!-\!2M^2_\Delta)^2\!-\!4M^4_\Delta)}\right)\right. \nn \\
&\times&\left.\left(\frac{(s_4\!-\!m^2_N)^2}{2W^2}\!-\!2m^2_\pi
\!-\!\frac{(2W^2M^2_\Delta\!-\!W^2(s_4\!-m\!^2_N))^2}
{2W^2((W^2\!-\!2M^2_\Delta)^2\!-\!4M^4_\Delta)}\right)\right]^{\frac12}\!\!.
\nonumber
 \ee
At $s_3=s_4$, Eq. (\ref{a6box})  reads
 \be \label{a6box-1}
s_{12}^{\rm box}=4m^2_\pi+\frac{W^2(2M^2_\Delta-s_3+m^2_N)^2}
{(W^2-2M^2_\Delta)^2-4M^4_\Delta}.
 \ee
Recall that in
(\ref{a6box}) and (\ref{a6box-1}) $M^2_\Delta$ is given by
\be
M^2_\Delta = m^2_\Delta -i m_\Delta\Gamma_\Delta .
\ee

\begin{figure}[h]
\leftline{\epsfig{file=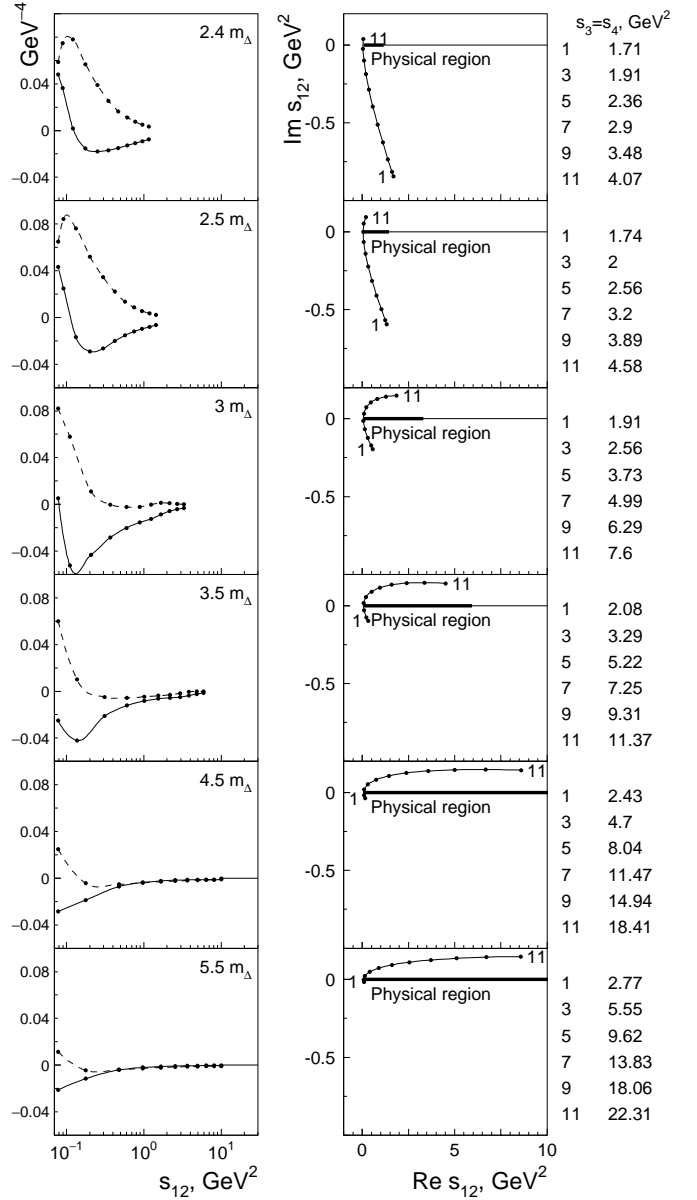,width=0.49\textwidth,clip=on}}
\caption{Box diagram amplitude as a function of $s_{12}$ under the constraint
(\ref{constraint}) (corresponding magnitudes of $s_3$ and $s_4$ are
shown in the right column. In the left columns, real and imaginary
parts of the amplitude are shown by solid and dashed curves
correspondingly. Initial energy, $W$, is shown on the top of each
panel. On the right columns singularity positions, $s_{12}^{\rm box}$,
Eq.(\ref{a6box}), are shown on the 2nd sheet of the complex-$s_{12}$
plane.} \label{fig_box_02}
 \end{figure}

\end{document}